\documentclass[a4paper,11pt]{article}
\pdfoutput=1 
\usepackage{jinstpub}
\usepackage{comment}
\usepackage{multirow}
\usepackage{booktabs}
\usepackage{parskip}
\usepackage{rotating}
\usepackage{dcolumn}
\usepackage{float}
\usepackage[nottoc,numbib]{tocbibind}
\usepackage{textcomp}
\usepackage{subfigure}
\usepackage{xspace}
\usepackage{tabularx}
\usepackage{adjustbox}
\usepackage[all, defaultlines=4]{nowidow}
\usepackage{wrapfig}
\usepackage{afterpage}
\usepackage{lineno}
\usepackage{array}
\usepackage{placeins}

\title{The SABRE South Technical Design Report Executive Summary}

 \affiliation[1]{School of Physics, The University of Melbourne, Parkville, VIC 3010, Australia}
\affiliation[2]{Department of Nuclear Physics and Accelerator Applications, The Australian National University, Canberra, ACT 2601, Australia}
\affiliation[3]{Department of Physics, The University of Adelaide, Adelaide, SA 5005, Australia}
\affiliation[4]{Centre for Astrophysics and Supercomputing, Swinburne University of Technology, Hawthorn, VIC 3122, Australia}
 \affiliation[5]{ARC Centre of Excellence for Dark Matter Particle Physics, Australia}
\affiliation[6]{Department of Physics, The University of Toronto, ON M5R 2M8, Canada}
 \affiliation[7]{School of Physics, The University of Sydney, NSW 2006 Camperdown, Sydney, Australia}
 \affiliation[8]{KEK, Oho, Tsukuba, Ibaraki 305-0801, Japan}
 \affiliation[9]{INFN Sezione di Milano, via Celoria 16, 20133 Milano, Italy}
 \affiliation[10]{INFN, Laboratori Nazionali del Gran Sasso, 67100 Assergi (L’Aquila), Italy}
 \affiliation[11]{INFN, Sezione di Roma, 00185 Rome, Italy}

\author[1,5]{E.~Barberio,}
\author[1,5]{T.~Baroncelli,}
\author[2,5]{V.~U.~Bashu,}
\author[2,5]{L.~J.~Bignell,}
\author[3,5,9]{I.~Bolognino,}
\author[4,5]{G.~Brooks}
\author[1,5]{S.S.~Chhun}
\author[2,5]{F.~Dastgiri,}
\author[10]{A.~Di~Giacinto,}
\author[11]{G.~D’Imperio,}
\author[4,5]{A.~R.~Duffy,}
\author[2,5]{M.~B.~Froehlich,}
\author[7,5]{T.~Fruth,}
\author[1,5]{G.~Fu,}
\author[3,5]{G.~C.~Hill,}
\author[1,5]{R.~S.~James,}
\author[3,5]{K.~Janssens,}
\author[7,5]{S.~Kapoor,}
\author[2,5]{G.~J.~Lane,}
\author[3,5]{K.~T.~Leaver,}
\author[11]{A.~Mariani,}
\author[3,5]{P.~McGee,}
\author[2,5]{L.~J.~McKie,}
\author[2,5,6]{P.~C.~McNamara,}
\author[1,5]{J.~McKenzie,}
\author[1,5]{W.~J.~D.~Melbourne,}
\author[1,5]{M.~Mews,}
\author[1,5]{G.~Milana,}
\author[1,5,b]{L.~J.~Milligan,}
\author[4,5]{J.~Mould,}
\author[11]{V.~Pettinacci,}
\author[1,5]{K.~J.~Rule,}
\author[4,5]{F.~Scutti,}
\author[2,5]{Z.~Slavkovsk\'{a},}
\author[1,5]{O.~Stanley,}
\author[2,5]{A.~E.~Stuchbery,}
\author[8]{B.~Suerfu,}
\author[1,5]{G.~N.~Taylor,}
\author[2,5]{D.~Tempra,}
\author[2,5]{T.~Tunningly,}
\author[1,5,a]{P.~Urquijo,}
\author[10]{C.~Vignoli,}
\author[3,5]{A.~G.~Williams,}
\author[1,5]{Y.~Xing,}
\author[1,5,6]{M.~J.~Zurowski}

\collaboration{The SABRE South Collaboration}

\collaboration{SABRE South Collaboration}
\emailAdd{sabre-contact@lists.unimelb.edu.au}
\emailAdd{(a) purquijo@unimelb.edu.au}
\emailAdd{(b) lmilligan@student.unimelb.edu.au}

\abstract{In this technical design report (TDR) executive summary we describe the SABRE South detector to be built at the Stawell Underground Physics Laboratory (SUPL). The SABRE South detector is designed to test the long-standing DAMA/LIBRA signal of an annually modulating rate consistent with dark matter by using the same target material. Located in the Southern Hemisphere, the detector is uniquely positioned to disentangle modulating seasonal effects. SABRE South uses seven ultra-high purity NaI(Tl) crystals (with a total target mass of either 35~kg or 50~kg), hermetically sealed in copper enclosures that are suspended within a liquid scintillator active veto. High quantum efficiency and low background Hamamatsu R11065 photomultiplier tubes are directly coupled to both ends of the crystal, and enclosed with the crystal in an oxygen free copper enclosure. The active veto system consists of 11.6~kL of linear alkylbenzene (LAB) doped with a mixture of fluorophores and contained in a steel vessel, which is instrumented with at least 18 Hamamatsu R5912 photomultipliers. The active veto tags key radiogenic backgrounds intrinsic to the crystals, such as ${^{40}}$K, and is expected to suppress the total background by 27\% in the 1--6~keV region of interest. In addition to the liquid scintillator veto, a muon veto is positioned above the detector shielding. This muon veto consists of eight EJ-200 scintillator modules, with Hamamatsu R13089 photomultipliers coupled to both ends. With an expected total background of 0.72~cpd/kg/keV, SABRE South can test the DAMA/LIBRA signal with 5$\sigma$ discovery or 3$\sigma$ exclusion after two years of data taking.}


\begin{document}
\setcounter{tocdepth}{2}
\maketitle
\flushbottom

   \section{Introduction}
\label{sec:intro}

Summarised in this document is the Technical Design Report for the SABRE (Sodium Iodide with Active Background REjection) South experiment. SABRE South is a NaI(Tl) dark matter direct detection experiment, to be situated underground in the Stawell Underground Physics Laboratory in the Southern Hemisphere, and aims to test the results from the DAMA/LIBRA experiment. SABRE South's three key sub-detectors are described, alongside other major components of the experiment design such as data acquisition, and shielding. The expected sensitivity of the detector to a DAMA/LIBRA-like signal is also discussed.

\subsection{Scientific goals}

A wide range of astrophysical evidence supports the existence of non-luminous matter known as dark matter (DM)~\cite{Freese:2008cz, Froborg:2020tdh, Rubin:1970, Zwicky:1937zza, Planck:2015fie}. Based on these observations, dark matter must have the following properties:
    it interacts gravitationally, but is electrically neutral and does not interact via the strong force;
    it is largely non-relativistic (``cold'');
    it is around five times more abundant than baryonic matter;
    it has a freeze-out time that allows for the observed structural evolution of the universe; and,
    it is stable on cosmological timescales.
Any observable signal from direct detection should have a distinct modulation signature due to the Earth's revolution around the Sun as it moves through the galactic dark matter halo~\cite{Froborg:2020tdh}. This modulation is unrelated to the Earth's seasons and  is identical in both hemispheres. To be compatible with DM, the modulation signal should satisfy the following conditions~\cite{FreeseAnnualMod, Froborg:2020tdh}:
    the modulation should have a period of one year;
    the modulation's phase should produce a peak at the start of June;
    for GeV-scale dark matter particle(s), the modulation should be present in a low energy range (keV-scale);
    the dark matter particle interacts rarely with matter, expecting only single hits with the target material;
    the modulation amplitude is not larger than $\sim$30\% of the total dark matter signal (the exact amount varies depending on the particle interaction model).

The primary motivation of SABRE (defined in Ref.~\cite{SABRE_POP}, which details the prototype detector) is to test a long-standing claim of a signal of dark matter interactions by another sodium iodide-based detector called DAMA/LIBRA, which has a significance of 13.7$\sigma$~\cite{Bernabei:2023gsv}.  If the signal observed by DAMA/LIBRA originates from a seasonally modulating background, SABRE South has the ability to identify or exclude the DAMA/LIBRA claim, due to its location in the Southern Hemisphere. SABRE South is the Southern Hemisphere component of the full SABRE concept, which also consists of the SABRE North experiment~\cite{SABRENorth} in the Northern Hemisphere. By using ultra-pure crystals with a background lower than those used in DAMA/LIBRA, and twin detectors in both hemispheres, SABRE aims to test the DAMA/LIBRA dark matter signal. SABRE South is capable of 5$\sigma$ CL exclusion after 3.1 years of data (see Sec.~\ref{sec:sensitivity}). SABRE South's sensitivity to WIMPs is calculated in Refs.~\cite{Zurowski_2021, sabre_background}, and has been compared to other NaI(Tl) experiments including DAMA/LIBRA.

The SABRE South detector is also capable of testing other models of dark matter and their interactions, as well as detecting supernova neutrinos out to the galactic centre, as secondary goals. Preliminary sensitivity studies have been performed for the detection of spin-dependent WIMPs via the Migdal effect~\cite{Migdal}, the detection of bosonic super WIMPs~\cite{BSWs}, and for the use of the SABRE South detector in a supernovae early warning system (SNEWS)~\cite{SNEWS:2020tbu}. Utilising the SABRE South liquid scintillator veto, we would expect to observe approximately 6 events from a typical core-collapse supernova event at 10~kpc, assuming a 3-fold coincidence requirement and the nominal number of PMTs. While the number of detection events would be smaller than other detectors in the SNEWS system, an additional experiment in the Southern Hemisphere, would be seen as valuable. Publications discussing the SABRE South detector's sensitivity to various dark matter models, and to supernovae neutrinos, are expected in the future.

\subsection{Dark Matter searches in NaI(Tl) detectors}
\label{ssec:dama}
\begin{figure}[htb]
    \centering
    \includegraphics[width=0.8\textwidth]{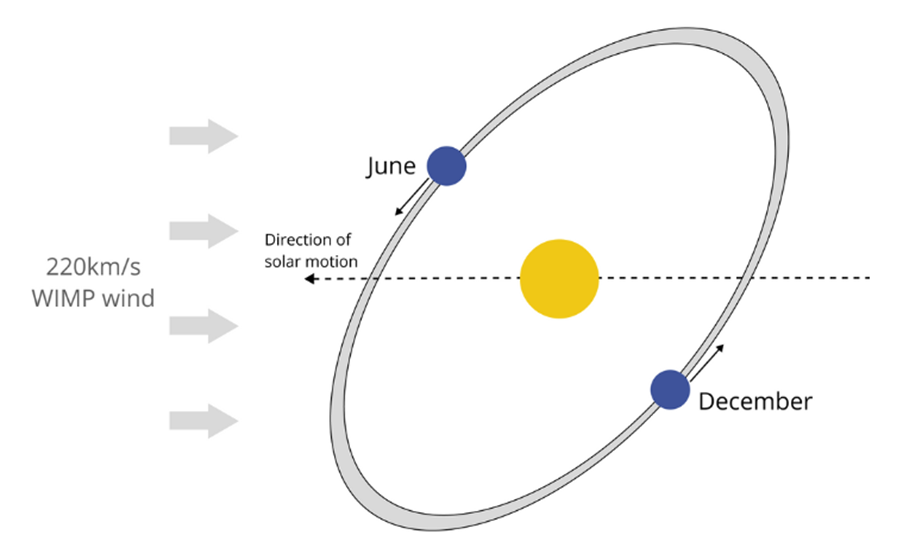}
    \caption{A diagrammatic depiction of the Earth's motion through the WIMP galactic halo (the WIMP wind effect), and the times of year at which the rate is at a maximum and at a minimum.}
    \label{fig:wimpwind}
\end{figure}

DAMA/LIBRA, like its predecessor DAMA/NaI, aims to observe the model-independent modulation of a dark matter interaction rate. The most recent iteration, DAMA/LIBRA-phase2 used 250 kg of NaI(Tl) over 8 years of data collection. Combining this with previous experiments, DAMA/LIBRA has a total exposure period of 2.86 tonne years. The interaction rate itself, integrated over a given energy window $k$, can be given by $$R_k(t) \approx R_{0,k} + S_{m,k} \cos(\omega(t - t_0)),$$ where $R_{0,k}$ is the constant rate, $S_{m,k}$ is the amplitude of the modulating component of the rate, $\omega$ is the angular frequency, and $t_0$ is the phase \cite{BERNABEI2020}. Figure~\ref{fig:wimpwind} depicts how the modulation is induced by the Earth's motion through the WIMP wind.

The DAMA collaboration has consistently observed a modulating signal over their background rate of 0.8~cpd/kg/keV~\cite{BERNABEI2020} (counts per day/kg/keV) throughout the DAMA/NaI, DAMA/LIBRA-phase1 and DAMA/LIBRA-phase2 experiments in the 2--6~keV region, which has a modulation amplitude of $0.01014 \pm 0.00074$~cpd/kg/keV with a phase and period of 142.4~$\pm$~4.2 days and 0.99834~$\pm$~0.00067~years respectively~\cite{bernabei2021}. 
In the 1--6~keV recoil energy range (reported for DAMA/LIBRA-phase2 only) this modulation occurs with an amplitude of $0.01058 \pm 0.00090$ cpd/kg/keV, period of $0.99882\pm0.00065$~years and phase of $144.5 \pm 5.1$~days. Recent preliminary results indicate that this modulation persists below the 1~keV threshold~\cite{Bernabei:2023gsv}.

The signal observed by DAMA/LIBRA is in direct tension with a wide variety of direct-detection dark matter experiments using different target materials~\cite{ParticleDataGroup:2024cfk}. Recently, there have also been studies suggesting that DAMA/LIBRA's background averaging technique (if performed over a time-dependent background) could have induced a modulating result~\cite{COSINEInducedMod}. However, after reproducing a best estimate of their background, we find that: (i) the reproduced background is too low for a modulation to be induced on the same scale as their reported signal, and (ii) the resulting induced modulation is opposite in phase to the modulation reported by DAMA/LIBRA~\cite{SABREInducedMod}.
Therefore, in recent years, other NaI(Tl)-based experiments have been developed to test this modulation using the same target as DAMA, such as COSINE~\cite{COSINE2024} and ANAIS~\cite{Coarasa:2024xec}.
We list the observed modulation from each of the operating experiments in Table~\ref{tab:global_mod}. Results in the 1--6~keV region ANAIS are known to be impacted by higher than expected background contributions that are unexplained \cite{Adhikari2021, Coarasa:2024xec}; a potential cause of this is photomultiplier noise~\cite{ANAISBDT}. Additionally, the signal efficiencies of the event selection criteria drop steeply as 1~keV is approached. For example, between 1--2~keV, the ANAIS detection efficiency drops to as low as 20\%, with the degree of background rejection not reported~\cite{Coarasa:2024xec}. For COSINE, the detection efficiency is around 40\% at 1~keV, with a sharp efficiency turn-on effect over the energy interval of 1--2~keV~\cite{COSINE-100LowThres}, which may lead to non-negligible systematic uncertainties. As a result, a comparison of results in the 2--6~keV region is likely more robust.
To definitively test DAMA/LIBRA, SABRE must be sensitive to a modulation of 0.01~cpd/kg/keV, which requires a total background of no more than 1 cpd/kg/keV, and ideally better than the DAMA value of 0.8 cpd/kg/keV~\cite{BERNABEI2020}.

\begin{table}[htb]
\centering
\caption{Annual modulation results reported by DAMA \cite{bernabei2021}, COSINE \cite{COSINE2024, Adhikari2021}, and ANAIS \cite{AnaisIDM} (ANAIS also reports updated results from their 3-year data \cite{Coarasa:2024xec}). We define dru (differential rate units) as cpd/kg/keV.
}
\label{tab:global_mod}
\begin{tabular}{@{}lllll@{}}
\toprule
Setup        & Mass (kg) & Bkg. (dru) & $S_m$ (1--6~keV) (dru) & $S_m$ (2--6~keV) (dru) \\ \midrule
DAMA/LIBRA-phase2           & 250 & 0.8 & ~0.0105 $\pm$     0.0009 &  0.0093 $\pm$ 0.0009   \\
COSINE         & 61.3 & 2.7&  ~0.0017  $\pm$     0.0029  & 0.0051 $\pm$ 0.0047      \\
ANAIS          & 112.5 & 3.2& ~0.0007 $\pm$    0.0025 &  0.0030 $\pm$ 0.0025  \\ \bottomrule
\end{tabular}
\end{table}

   \section{The Stawell Underground Physics Laboratory}
\label{sec:supl}

\begin{figure}[htb]
    \centering
    \includegraphics[width=\textwidth]{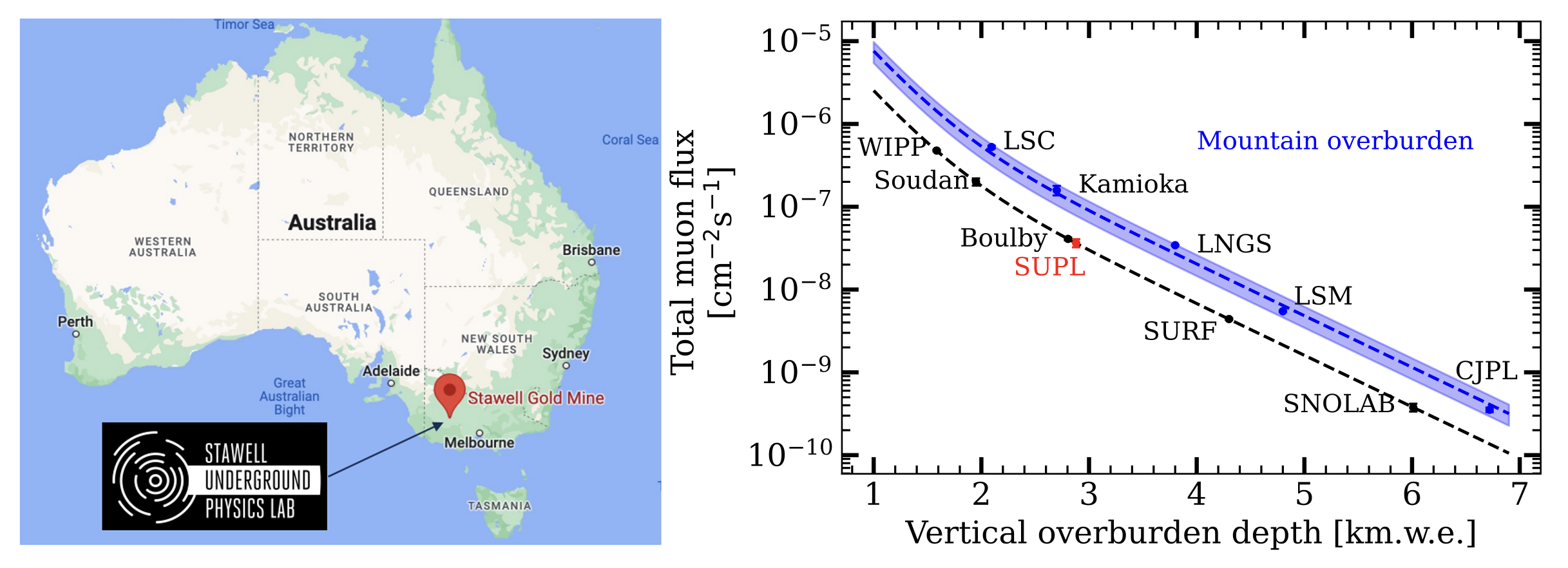}
    \caption{Left: location of SUPL. Right: muon flux at a number of underground laboratories as a function of water equivalent depth. The point for SUPL is based on simulated results using GEANT4~\cite{GEANT4} and MUTE~\cite{MUTE}. The points denote flux measurements in each laboratory. The dashed lines are fits to the data points according to the method laid out in Ref.~\cite{JNEMuonsDepth}. The black line and points represent labs with a flat overburden, and the blue line and points represent labs with mountainous overburden (with the blue region denoting the error on the fit). Not included in the figure is the data for Y2L: publicly available data indicates that Y2L has depth of about 2.4~km of water equivalent with mountainous overburden~\cite{Kim:2014toa}.}
    \label{fig:supl}
\end{figure}

The SABRE South detector is hosted at the Stawell Underground Physics Laboratory (SUPL), which is 1025 m underground in Victoria, Australia. SUPL is the first deep underground laboratory in the southern hemisphere. The laboratory has a flat overburden of basalt, providing approximately 2900~m of water-equivalent shielding, and a muon flux of $\sim 3 \times 10^{-8}$~cm$^{-2}$s$^{-1}$ (Fig.~\ref{fig:supl}).

The construction of SUPL was completed in 2023, with the first access for SABRE South installation activities in early 2024. The assembly and commissioning of the main detector components in SUPL is currently scheduled for 2025, after which data collection will begin. SUPL has helical drive access through mining tunnels and no vertical access shafts, which limits the size and weight of detector components. The Stawell Gold Mine remains an active gold mine; however, SUPL is located in a disused portion of the mine.

The walls of SUPL, covered with a pre-screened low background concrete mixture, were finished with two coats of Tekflex sealant to inhibit radon flow into the lab through the walls. A radon-reduction system was part of the original SUPL design; however, due to financial constraints, the system has not yet been implemented. The airflow into the SUPL air-conditioning system arrives through approximately 9~km of decline tunnels, drawn down by large fans pumping air at high rate through separate vent shafts. This air and other inlet air pick up radon gas from the untreated walls along the way. A compressed air line carrying air from the surface to the depth of SUPL will be used to feed an over-pressure of low-radon air into clean-tent environments within SUPL during assembly of the detector. 

The measured radon activity in the laboratory is $415 \pm 5$~Bq/m$^{3}$, which requires special precautions during the assembly and design of the experiment. This includes the need to avoid radon decay daughter products, such as $^{210}$Pb, from depositing (``plating'') onto sensitive parts of the detectors and related equipment.
 
\begin{figure}[htb]
    \centering
    \includegraphics[width=0.5\textwidth]{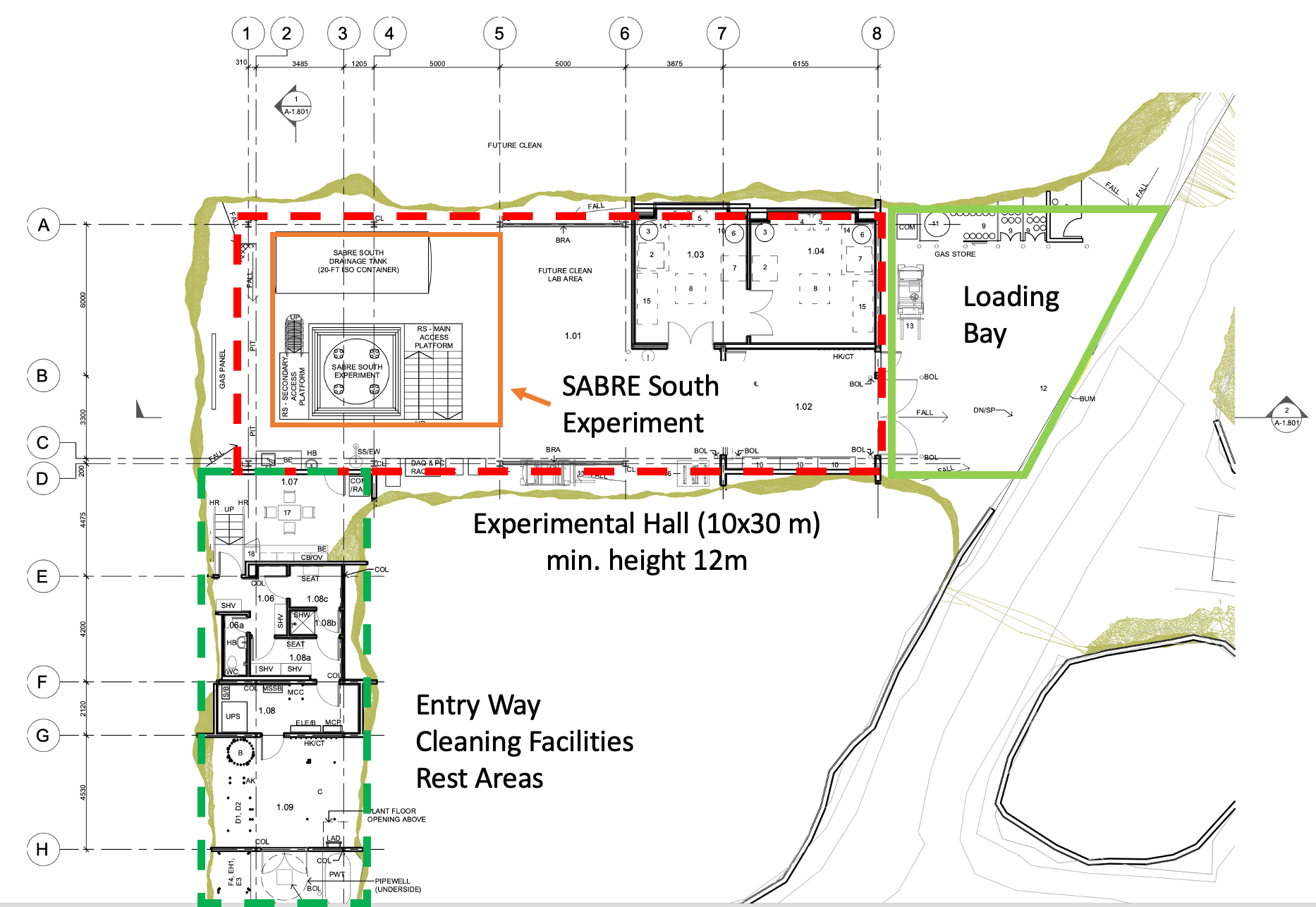}
    \includegraphics[width=0.45\textwidth]{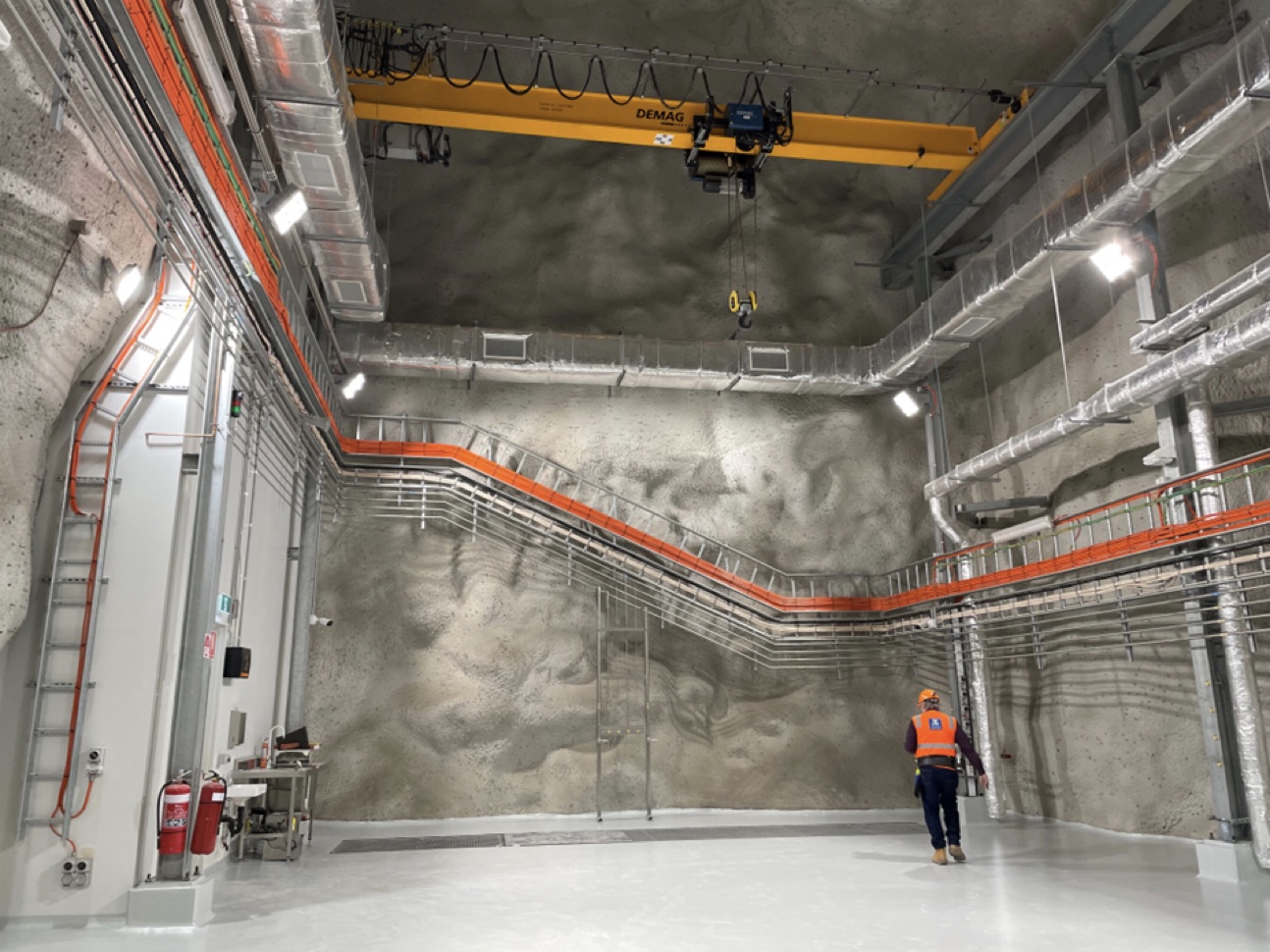}
    \caption{Left: Schematic of SUPL including the main experimental hall (top left), low background counting (HPGe) facilities (top middle), loading bay (top right), and the entry way with cleaning facilities (bottom left). Right: Photograph of SUPL. SABRE South will be located near the centre of the image. The door on the left leads to the rest areas and entry way.}
    \label{fig:supl-schem}
\end{figure}

A schematic of the laboratory (including the location of SABRE South) is shown in Fig.~\ref{fig:supl-schem}, alongside a photograph of the laboratory. The main hall has floor dimensions of 10~m by 26~m and a maximum height of 12~m. On the side of the main hall are the cleaning facilities and entry way.

   \section{The SABRE South concept}
\label{sec:exp}

To definitively test the DAMA/LIBRA anomaly, it is essential to use the same target material and achieve an ultra-low background to detect the reported modulation amplitude of 0.01 cpd/kg/keV. Since the primary contributor to the total background is the radiogenic background in the crystals, the SABRE South experiment aims to achieve background levels comparable to those of DAMA/LIBRA by ensuring high NaI(Tl) radiopurity. Additionally, by employing a background veto system—comprising the liquid scintillator detector and the muon detector—external background and residual radioimpurities can be tagged and removed~\cite{Calaprice:2021yml}.

The test crystals grown for the SABRE collaboration are very competitive with other NaI(Tl) crystals in terms of radiopurity~\cite{Amare2019, Adhikari, Suerfu2019, antonello2021}. SABRE South's location in the Southern Hemisphere also offers a unique advantage by providing data that can disentangle seasonal effects.

\subsection{Radiopurity of NaI(Tl) detectors}
\label{sec:crys-dev}

\begin{figure}[htb]
    \centering
    \includegraphics[width=0.6\textwidth]{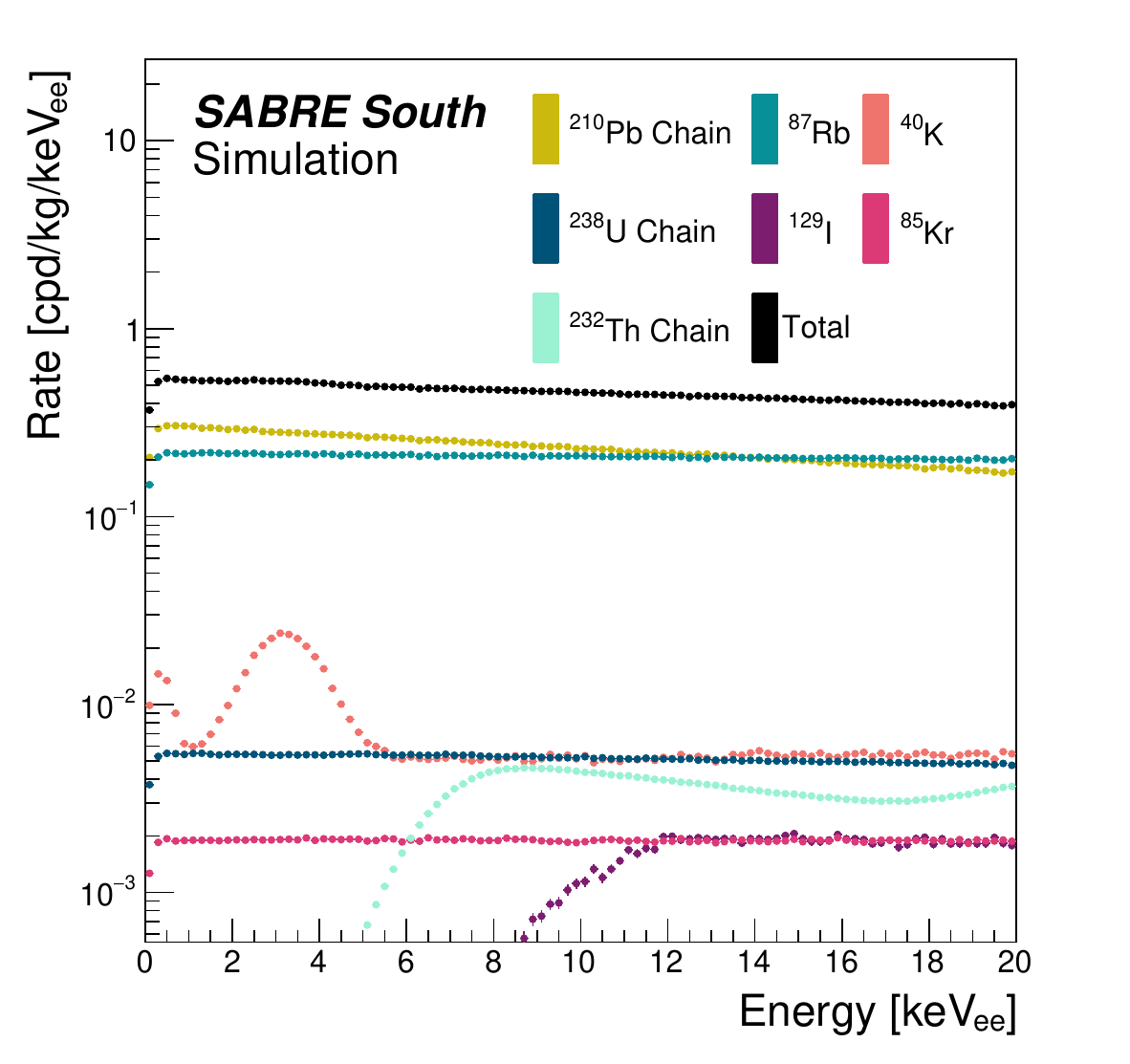}
    \caption{The crystal background rate from intrinsic contamination at SABRE South for 50~kg of NaI(Tl) crystals~\cite{sabre_background}. The units keV$_{\rm ee}$ refer to electron equivalent energy. These spectra are based on measurements of the SABRE Proof-of-Principle crystal~\cite{antonello2021}. Note: The y-axis on this does not extend down to zero, and the spectra for $^{129}$I and $^{232}$Th continue below the lower y-axis limit. Details on their total background rate below this limit can be found in Ref.~\cite{sabre_background}.}
    \label{fig:sabre_crys_bkg}
\end{figure}

SABRE South and SABRE North are designed such that radiogenic background intrinsic to the crystals dominate the total background, as demonstrated through simulations~\cite{sabre_background, SABRE_MC} and indicated in Refs.~\cite{Adhikari, Amare2019}. This design highlights that radiopure crystals are the key factor in achieving an experiment capable of reproducing the DAMA/LIBRA sensitivity.  The purity levels of the DAMA, ANAIS, COSINE, and SABRE crystals are listed in Table~\ref{tab:competition}, with the SABRE South crystal radiogenic background spectrum (Fig.~\ref{fig:sabre_crys_bkg}) taken from the measurements in Ref.~\cite{antonello2021}, which utilise the SABRE Proof-of-Principal crystal. The veto efficiency is applied. 
 The SABRE crystals have a  $^{40}$K contamination better than that of DAMA/LIBRA and much lower than than the other existing NaI(Tl) experiments (COSINE and ANAIS). Another important contaminant is $^{210}$Pb that can accidentally enter the crystal during powder production and handling, or in the growth procedure, due to exposure to $^{222}$Rn~\cite{antonello2021, Adhikari, Yu:2020ntl}. The background spectra in Fig.~\ref{fig:sabre_crys_bkg}, including the $^{210}$Pb spectrum, are based on measurements reported in Ref.~\cite{antonello2021}. The $^{210}$Pb background spectrum derives from concentrations measured after the crystal was chemically polished with semiconductor-grade ethanol/isopropyl alcohol.
 
 As shown in Ref.~\cite{antonello2021}, SABRE crystals have the lowest NaI(Tl) background worldwide, enabling SABRE to achieve the 0.72~cpd/kg/keV total background reported in Ref.~\cite{sabre_background}. Further details on NaI(Tl) background modelling for SABRE South can be found in Ref.~\cite{sabre_background}, and crystal characterisation is discussed in Ref.~\cite{antonello2021}.

SABRE South is considering two different final scenarios for crystal growth: one with a total mass of approximately 50 kg (7$\times$7 kg crystals grown using the Bridgman method by the Shanghai Institute of Ceramics, Chinese Academy of Sciences (SICCAS)), and another with a total mass of 35 kg (7$\times$5 kg crystals grown using zone refining followed by the Bridgman method by Radiation Monitoring Devices (RMD), Boston). References to a total active mass of 35-50 kg reflect these different scenarios. In both cases the Bridgman method is chosen, as opposed to the Kyropoulos technique, as the molten raw material can be completely sealed inside the crucible during crystal growth, reducing the risk of contamination.

The total background and sensitivity for the 50 kg scenario are discussed in Sec.~\ref{sec:sensitivity}, where the total background of 0.72~cpd/kg/keV (with the radiogenic crystal background of 0.52~cpd/kg/keV~\cite{sabre_background} after the veto is applied) allows for a 5$\sigma$ exclusion in three years and a 5$\sigma$ discovery in two years. The 35 kg scenario requires stricter background limits to maintain the same sensitivity. For the 35 kg case, to achieve a 5$\sigma$ discovery or 3$\sigma$ exclusion, a total background of less than 0.5~cpd/kg/keV is needed. Assuming the same NaI(Tl) cosmogenic (cosmogenic refers to activation from cosmic rays) and extrinsic contributions to the total background, this means the NaI(Tl) radiogenic background must be less than 0.3~cpd/kg/keV~\cite{sabre_background}, which is achievable using the zone refining technique~\cite{SABRENorth}.

\begin{table}[ht]
\centering
\caption{Purity levels of NaI(Tl) crystals of various experiments.}
\label{tab:competition}
\begin{tabular}{@{}lcccc@{}}
\toprule
~Experiment & $^{\rm nat}$K (ppb) & $^{238}$U (ppt) & $^{232}$Th (ppt) & $^{210}$Pb (mBq/kg) \\ \midrule
~DAMA/LIBRA \cite{Bernabei2008} & $13$ & $0.7-10$ & $0.5-7.5$ & $(5-30)\times 10^{-3}$ \\
~ANAIS-112 \cite{Amare2019} & $31$ & $<0.81$ & $0.36$ & $1.53$ \\
~COSINE-100 \cite{Adhikari} & $35.1$ & $<0.12$ & $<2.4$ & $1.74$ \\
~SABRE (NaI-033) \cite{Suerfu2019, Tomei:2022qes} & $4.3$ & $0.4$ & $0.2$ & $0.5$ \\ \bottomrule
\end{tabular}
\end{table}

\subsection{SABRE South detector}
\label{ssec:detector}

\begin{figure}[htb]
    \centering
    \includegraphics[width=0.7\textwidth]{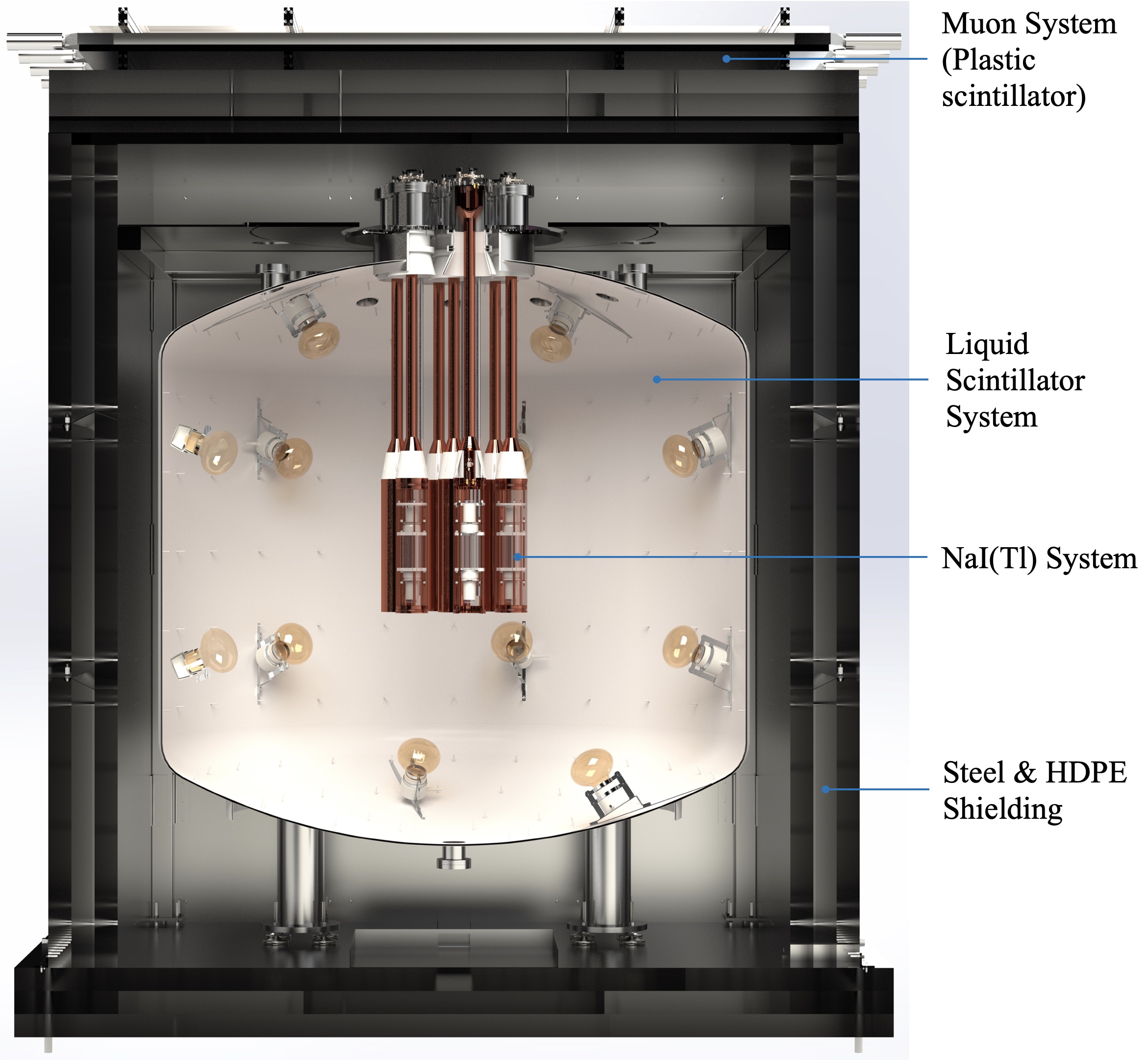}
    \includegraphics[width=0.7\textwidth]{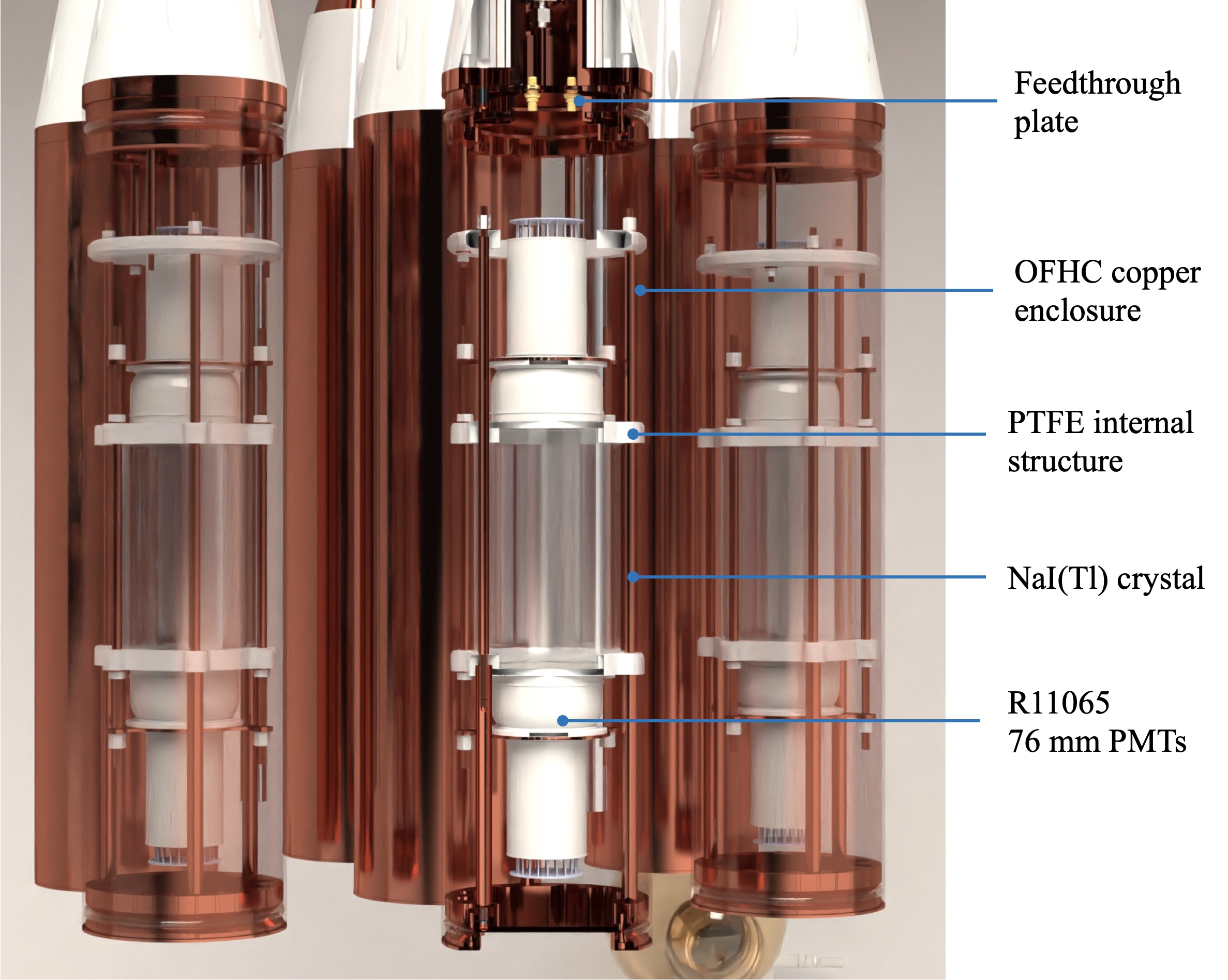}
    \caption{Top: Schematic view of the full SABRE South detector. Bottom: Schematic view of the crystal detector system where OFHC copper refers to oxygen free high thermal conductivity copper.}
    \label{fig:sabre-schem}
\end{figure}

The SABRE South experiment is designed to detect dark matter with a 35-50 kg NaI(Tl) target and an active background rejection system.
The apparatus is made up of three sub-detector systems: (i) the NaI(Tl) crystal detector array, (ii) the linear alkylbenzene (LAB) liquid scintillator veto system, and (iii) the plastic scintillator muon veto system. Together, the liquid scintillator and muon detectors act as an active veto, rejecting events from intrinsic and extrinsic background sources. These components are shown in Fig.~\ref{fig:sabre-schem} and are detailed in the following sections.

   \section{Crystal modules}
\label{sec:crys_modules}

The NaI(Tl) crystal detector system is the core of the SABRE South detector, and significant effort has been put into photomultiplier tube (PMT) characterisation, crystal growth, enclosure design, and insertion system design. The crystals are grown from Merck (formerly Sigma-Aldrich) Astrograde NaI powder, which has potassium contamination below 10~ppb \cite{suerfu-thesis}. Upon arrival at SUPL, they are stored underground and assembled in a nitrogen-flushed assembly glove box, housed in a radon- and dust-reduced space.

SABRE collaborates with SICCAS (China) and RMD (USA) to produce ultra-pure NaI(Tl) crystals. RMD has produced several prototypes that were tested at LNGS (Laboratori Nazionali del Gran Sasso) for optical and radiopurity properties. SICCAS is currently growing additional test crystals. Low levels of natural radiation within the NaI powder. These can produce energy depositions in the region of interest (1--6~keV) that obscure the DM signal. The isotopes that form a significant background near the region of interest are $^{40}$K and $^{210}$Pb, alongside others depicted in Fig.~\ref{fig:sabre_crys_bkg}. Note that the $^{87}$Rb contribution is from an upper limit measurement and is expected to be significantly lower in practice. We plan to measure the quenching factors (QFs) of the crystals, using the tips and tails of each crystal ingot, as the QFs may vary from crystal to crystal. SABRE South has performed a QF measurement on a test NaI(Tl) crystal in Ref.~\cite{BignellQF}, which details the procedure that we plan to follow for later measurements. 

Each crystal is coupled with optical grease to two 76 mm-diameter Hamamatsu R11065 PMTs, and is enclosed in an oxygen-free, high-thermal-conductivity (OFHC) copper enclosure. The sides of the crystals are wrapped in PTFE reflector to ensure optimal light collection. The PTFE can contain $^{210}$Pb contaminants. The contribution of this radioimpurity to the background depends on whether it is on the surface or in the bulk of the material. Deposits close to or on the surface will lead to higher background levels~\cite{Tomei:2022qes}. Improved simulations of light collection efficiency are being developed and will be compared with upcoming test measurements at SUPL. These crystal enclosures are then submerged in the SABRE South vessel, suspended by copper conduits that house the cables from the PMTs inside the vessel. The crystals are placed vertically rather than horizontally to ensure that they do not slide against the PMTs under gravity, and for practicality, as a smaller opening in the vessel is required for insertion.  They have been fabricated at the Australian National University and have been designed to be leak-proof when immersed in LAB for the duration of the experiment. To ensure a moisture-, oxygen-, and radon-free volume, the enclosures are flushed at a low flow rate with high-purity nitrogen, the handling and control of which is managed by a bespoke gas handling system (see Sec.~\ref{sec:GHS}).

After assembly, each crystal detector will undergo preliminary testing in a low-radon, nitrogen purged, shielded environment in SUPL, to ensure good performance and make initial measurements of the intrinsic (radiogenic and cosmogenic) background. After insertion into the veto vessel, in-situ characterisation and full background measurements will be performed. A summary of the components is given in Table~\ref{tab:sabre-specs-nai}.

\begin{table}[htb]
\centering
\caption{Components of SABRE South NaI(Tl) detector system including their specifications.}
\label{tab:sabre-specs-nai}
\begin{tabularx}{0.99\textwidth}{l l}
\toprule
\bf{Component}    & \bf{Specifications}   \\ \midrule
NaI(Tl) crystals  & 7 $\times$ 5--7 kg - length $\sim$15--20~cm, dia.~$\sim$10~cm   \\
Photomultipliers  & Hamamatsu R11065, 76 mm dia., 2 per crystal, QE $>$ 30\%      \\
Readout           & CAEN V1730 digitiser, 500~MS/s sample rate, 0.12~mV resolution\\ 
Enclosure         & OFHC copper shell and internal structure with PTFE (Teflon) components,\\
                  & sealed with viton O-ring   \\

\midrule
\end{tabularx}
\end{table}

Following earlier collaboration with Hamamatsu on the development of an ultra-low radioactive background in a PMT, SABRE South has procured the full set of 14 R11065 76~mm PMTs with $\ge30\%$ quantum efficiency. The single-photon response, dark rate, and afterpulse rate of each PMT are being characterised at The University of Melbourne prior to installation. 

\subsection{Enclosure design}
\begin{figure}[htb]
    \centering
    \includegraphics[width=0.75\textwidth]{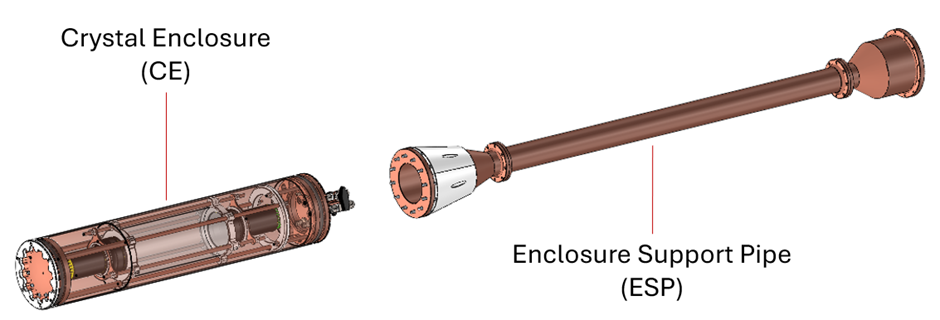}
    \includegraphics[width=0.75\textwidth]{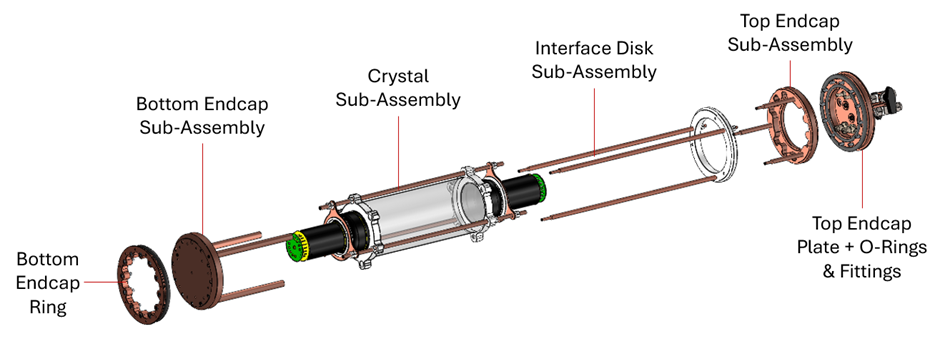}
    \caption{A render of a crystal enclosure, with the crystal and PMTs depicted. The enclosure will attach to a copper conduit/support pipe, also pictured. An exploded render of the crystal enclosure is also shown.}
    \label{fig:CEFullRend}
\end{figure}

SABRE South supports seven crystal detector modules. The enclosures are designed to ensure the aforementioned contaminant-free environment during handling, assembly, and insertion of the NaI(Tl) detector modules. A render of a crystal enclosure, with the associated conduit/support pipe, is shown in Fig.~\ref{fig:CEFullRend}. Each enclosure is suspended by an enclosure support pipe that is connected to the top endcap in the top flange of the veto vessel. It carries the PMT signal and high voltage (HV) connections, gas lines, and other sensor cables, while keeping them isolated from the liquid scintillator.

The enclosures must meet a strict set of requirements. The leak rate across the sealing surfaces can not exceed $1\times10^{-6}$~mbar~L/s. They are to be immersed in the LAB liquid scintillator for at least 5 years without degradation of the liquid scintillator or ingress of the scintillator into the enclosure. Long-term material compatibility studies performed by SABRE South have confirmed that all materials in direct contact with the liquid scintillator are compatible and are not expected to break down. The enclosures must provide a minimal contribution to the experimental background. The material between the crystals and the liquid scintillator must be minimised to achieve optimal veto efficiency. The dimensions of the enclosures must comply with the restrictions imposed by the crystal insertion system and the veto vessel top flange. The crystal dimensions are expected to vary slightly in size, so the enclosures must be adaptable enough to accommodate a range of sizes. The design allows a height of up to 250~mm, and a diameter limit of approximately 110~mm, corresponding to an 8.72~kg crystal. Finally, the assembly of the modules is performed inside a glovebox, so the design must accommodate assembly with a restricted range of motion.

The enclosures are made entirely of OFHC copper (C0100) and PTFE. The OFHC copper components are extremely radiopure while being commercially available, and were sourced from Thyssenkrupp in the United States. The copper and PTFE components are machined at the Australian National University (ANU). The copper surfaces are chemically etched to reduce $^{210}$Pb plate-out and other radioactive impurities, and the PTFE is vacuum baked to reduce out-gassing.
The outer shell of the crystal enclosure has an outer diameter of 143~mm, a maximum wall thickness of 3~mm and a length of 670~mm. The hermetic seals of the enclosure are achieved using `V-grooves' that compress O-rings with a 45$^\circ$ chamber in the endcap plates against the copper shell.

\begin{figure}[htb]
    \centering
    \includegraphics[width=0.7\textwidth]{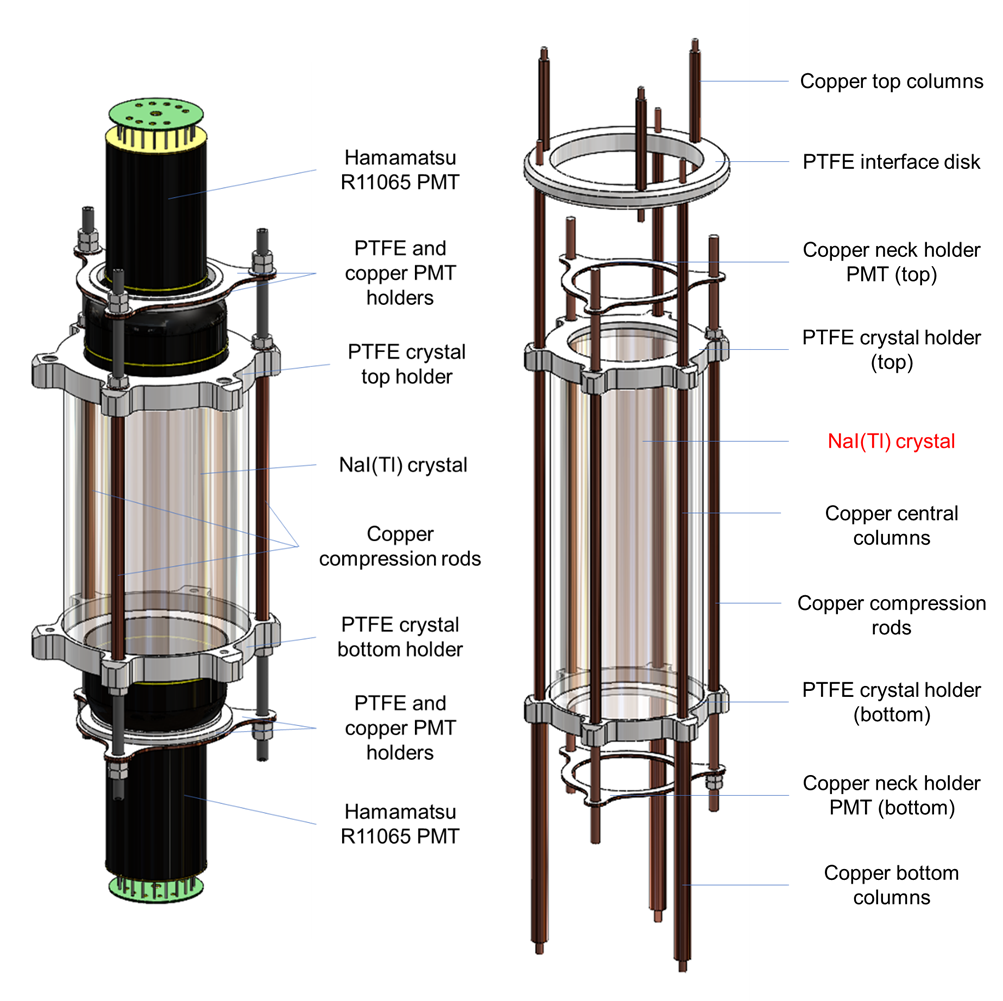}
    \caption{A render of the full internal subassembly supporting the NaI(Tl) crystal and coupled PMTs, with components labelled.}
    \label{fig:CrysSubAssemb}
\end{figure}

Shown and fully labelled in Fig.~\ref{fig:CrysSubAssemb}, the internal crystal sub-assembly is designed to mechanically support the NaI(Tl) crystal and the coupled PMTs within the copper shell, ensuring that coupling is maintained during experimental run time and through the later stages of assembly. The crystal is held with two PTFE holders that sit above and below the crystal with a cutout for the PMTs. These are connected with three OFHC-copper compression rods that provide mechanical support for the crystal-PMT coupling. Each PMT is held in place with a copper and PTFE holder and directly coupled to the crystal with optical grease, which is connected to the compression rods. This design provides an evenly distributed pressure on the PMTs and the optical grease, which ensures good optical coupling to the NaI(Tl) crystals without causing excessive stress on the PMTs. The compression rods are 7~mm in diameter with partial threading to attach the PMT holders, to allow for variations in crystal height.

The crystal sub-assembly is connected to the enclosure endcaps with three copper columns that, when assembled, provide the mechanical support for the sub-assembly, shown in Fig.~\ref{fig:CEFullRend}. As the crystal growth process may lead to crystals of differing dimensions, the crystal enclosure has been designed to accommodate variations both in crystal height and diameter. Consequently, there are components that are either independent of or dependent on the size of the crystal. The main components dependent on size are the compression rods, copper support columns, and the PTFE crystal holders. These are made to the crystal dimensions and are designed to align the crystal and veto vessel midplanes. Crystals with a height less than the limit will have more clearance at each end. The remaining components are independent of crystal size and allow for modular manufacturing and assembly prior to the crystal growth being finalised. 

A prototype crystal enclosure has been constructed with this design, and was used to conduct assembly tests to confirm the suitability of the design. In addition, a top endcap was manufactured to conduct leak testing using a helium leak checker, which will be repeated for each manufactured enclosure. The crystal enclosure components independent of crystal size have since been manufactured.

The crystal enclosures are assembled in a glove box from Xiamen LITH Machine Limited. It has a main chamber 1800~mm in length with four workstations, and an 800~mm ante-chamber for loading the crystals and components for assembly. The glove box is constructed from 3~mm stainless steel, and has 8~mm-thick plexiglass windows. Organic solvents can be used within the glove box for polishing of the crystals, should they be required. The ante-chamber can be pumped under vacuum and backfilled with N$_2$, and the main chamber operates with a constant flow of N$_2$. The expected concentration of O$_2$ and H$_2$O is at the ppm level.

\subsection{R11065 PMTs}

\begin{figure}[htb]
    \centering
    \includegraphics[width=0.425\textwidth]{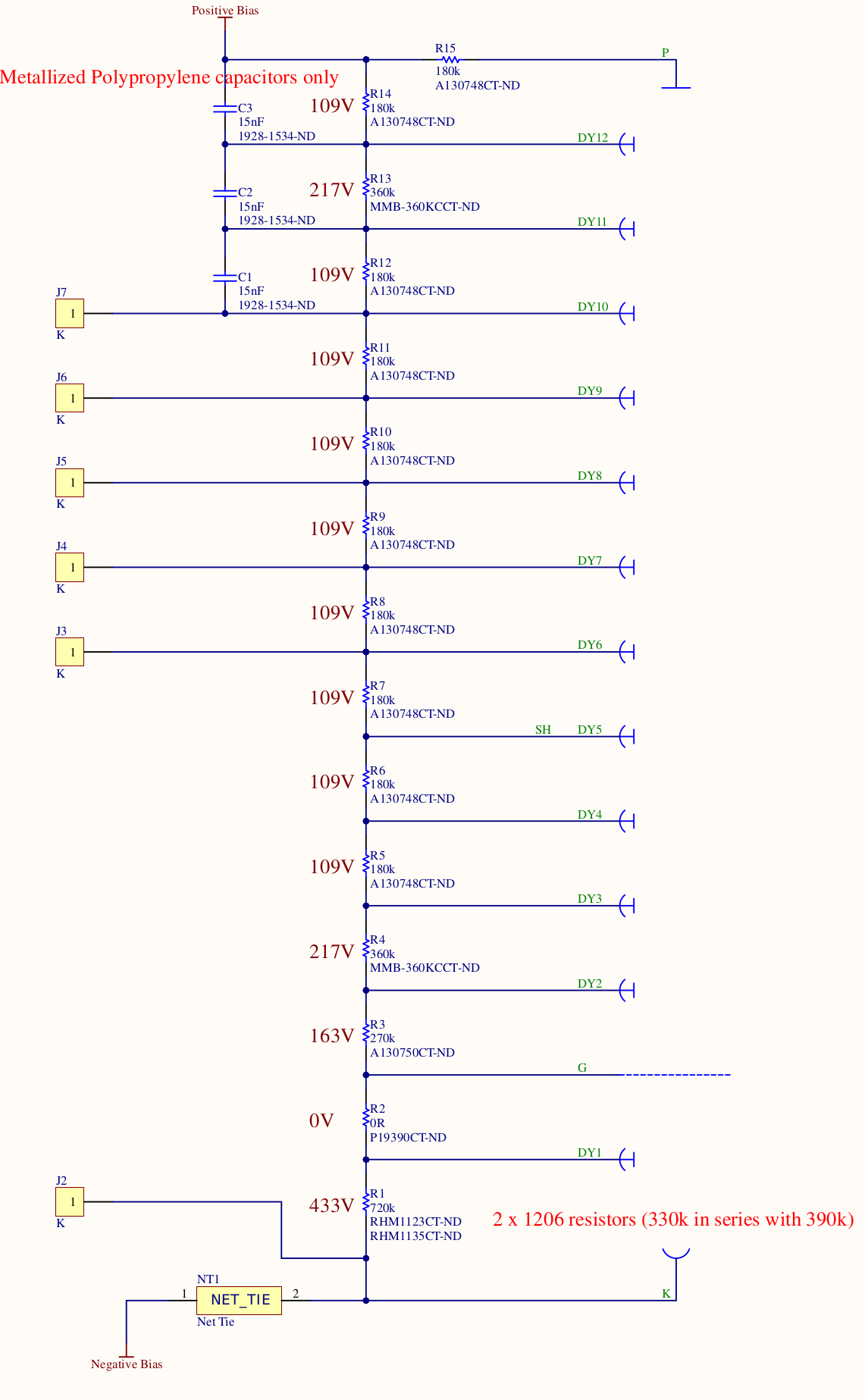}
    \includegraphics[width=0.5\textwidth]{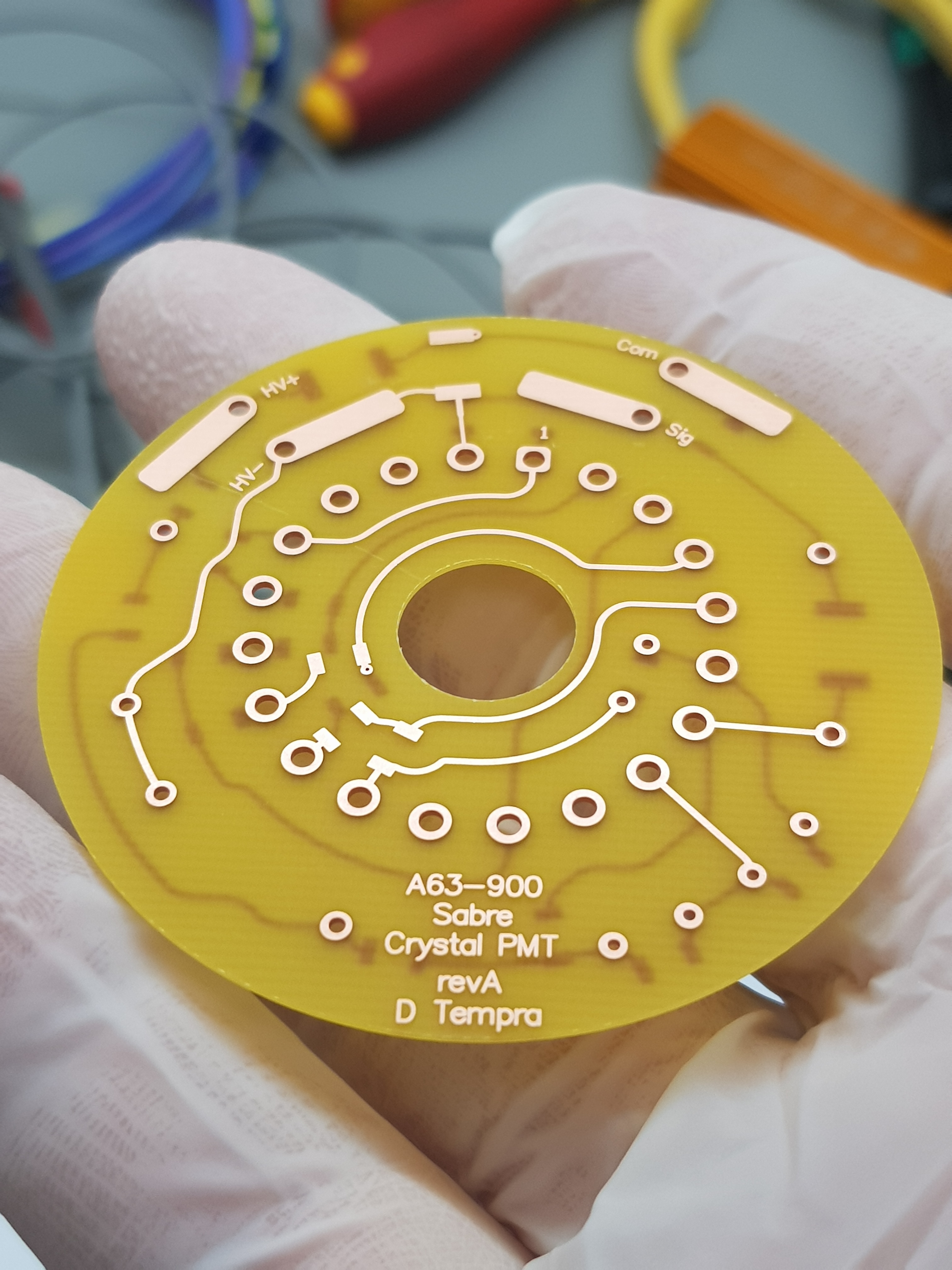}
    \caption{Left: The circuit diagram for the final R11065 base design. Right: One of the PCBs to be used for the base circuit, which will be attached to an R11065 PMT.}
    \label{fig:CrysPMTBase}
\end{figure}

The R11065 PMTs were chosen for their high sensitivity to single photons, low noise rates, and low radioactive background. Only PMTs with quantum efficiency (QE) $>30\%$ were procured, so to achieve a high single-photon detection efficiency. Custom PMT voltage dividers are used to minimise their contribution to the background. These bases will follow the resistance ratios recommended by Hamamatsu for this type of PMT. SABRE South has decided on a design that will use a positively biased voltage divider. The circuit diagram and an image of the PCB for the voltage divider design are shown in Fig.~\ref{fig:CrysPMTBase}. The final design is expected to use small ceramic capacitors. The nominal design does not include an onboard or offboard preamplifier. Preamplifiers are being considered for a future upgrade, but would not be installed onboard.

The pre-calibration program for the R11065 (crystal detector) PMTs focuses on confirming their single-photon detection efficiency, characterising their single-photo electron (PE) response, and characterising the noise as a function of temperature and voltage. These parameters are used as an early calibration input to SABRE South data analysis and as an input into the digitisation stage of detector simulations. The requirements for a quantum efficiency larger than $30 \%$ (at a wavelength of 420~nm) and excellent signal-to-background discrimination for single photons are driven by the light yield requirements (of $\sim$~12--14~PE/keV) on the crystal detectors to achieve a threshold of 1--2 keV. Several essential measurements will be completed for all PMTs under strict cleanliness protocols. These are: relative quantum efficiency, gain, dark rate, and afterpulse rate at nominal voltage. Further characterisation of single-photon response and noise, including temperature and bias voltage dependence, is performed on a limited set of PMTs to limit overall exposure. As SABRE operates at room temperature, dark counts due to thermionic emission and temperature dependence of this dark rate are essential factors that are more easily measured in a dedicated setup than in-situ. The pre-calibration datasets  allow for the development of data analysis techniques for discrimination of thermionic emission from low-energy events in the crystals, near the 1--2 keV region. 
There have previously been reports \cite{Barrow_2017,Akimov:2015alu,Akimov_2016} of light emission likely due to anode and dynode glow following a sufficiently intense light signal in the R11065 and the related R11410 PMT types. The R11410 and R11065 PMTs  differ only in the choice of photocathode material. Such light emission could introduce an additional background in SABRE and therefore needs to be characterised if present.

\subsection{Cleanliness protocols}

All components of the crystal enclosures are cleaned and vacuum baked prior to final assembly to ensure that there is no moisture remaining on the components. The crystal PMTs, cables, and bases are cleaned with lint-free isopropyl alcohol (IPA) soaked cleanroom wipes and a high pressure nitrogen gun to remove any debris. To prevent radon plate-out, all crystal detector module components are protected in two-layered radon bags for storage between cleaning and their installation in the final detector modules.  The assembly of the base/cables and their mounting on the PMTs is performed in an ISO6 cleanroom environment. Assembly of the crystal modules is performed in a nitrogen flushed glove box. 

\subsection{Crystal Insertion System}

\begin{figure}[htb]
    \centering
    \includegraphics[width=0.85\textwidth]{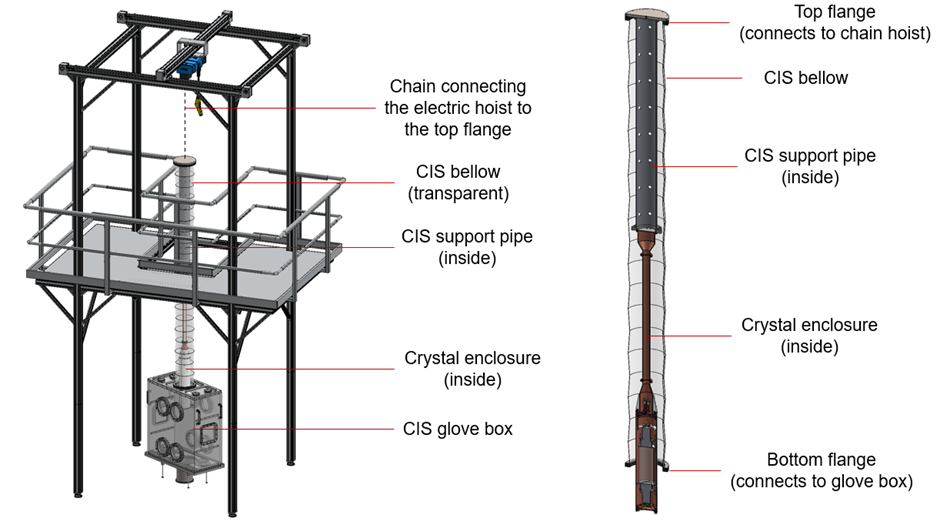}
    \caption{A render of the CIS, including the support frame and glovebox with associated compressible bellows.}
    \label{fig:CISRender}
\end{figure}

The crystal detectors are installed in the veto vessel using a custom-made crystal insertion system (CIS) 
that is positioned above the vessel during installation. The CIS is a 4.5~m-tall aluminium frame with a computer-controlled 2-D stage for positioning a mobile gantry crane directly above the veto vessel ports, and a nitrogen-flushed glovebox to insert the crystal detectors into the vessel. A complete annotated render of the system is shown in Fig.~\ref{fig:CISRender}.  

The CIS has been developed for the insertion of the crystal detectors under both `wet' (veto vessel filled with liquid scintillator) and `dry' (before filling with liquid scintillator) conditions. The requirement of supporting wet insertion is to allow for staged installation of the crystal modules and to ensure that any crystal module can be accessed during operation of the experiment in the event of a technical issue or upgrade.

The main components of the system are the CIS glove box and the support frame, depicted in Fig.~\ref{fig:CISRender}. The CIS glove box is a custom-made PMMA acrylic glove box manufactured by Palazzi SRL, composed of a single main chamber with a compressible bellows attached to the top. The main glove box chamber has internal dimensions of 950~mm$\times$450~mm$\times$965~mm (0.366~m$^3$), with four pairs of gloves for module handling. During use, the CIS glove box is supported by a composite steel-aluminium structure mounted on the veto vessel top flange.

The compressible bellows is used to maintain a sealed environment throughout the insertion process, with up to 2000~mm of vertical travel, and is operated with the computer-controlled electric hoist. A gas handling system is used to keep the glove box free of LAB vapour contamination when the movable bellows is in use.
The CIS frame provides support for the glove box and hoist and is mounted on top of the work platform during use. It is constructed from aluminium sections and has been assembled.

The process of insertion involves inserting the entire enclosure and conduit within the compressible bellows, which is attached to the chain hoist at the top and to the CIS glove box at the bottom. The CIS glove box is mounted on the relevant crystal flange. The entire system is nitrogen flushed. The chain hoist lowers the enclosure into the vessel through the flange and is guided by hand through the glove box. The bellows compresses as the chain hoist lowers, ensuring any residual LAB vapour is forced back into the vessel. Upon full insertion, the enclosure is attached to the crystal flange, thus sealing the flange. This process is repeated for each enclosure.

\subsection{Gas Handling System}
\label{sec:GHS}

The NaI(Tl) crystals are hygroscopic and must be kept in a dry environment both when handled during storage and assembly, and once installed in the experiment. The assembly of the enclosures is carried out in a nitrogren-filled glove box, and they are sealed for installation. Once installed in the LS vessel, the enclosures are flushed with a low flow rate of high-purity nitrogen (HPN), ensuring a moisture-, oxygen-, and radon-free volume. Individual enclosures also include exhaust flows for sampling and monitoring of volatile organic compounds, water, and oxygen. Furthermore, within the veto vessel the LAB must not be allowed to come into contact with O$_2$, and therefore requires HPN blanketing.

\begin{figure}[htb]
    \centering
      \rotatebox{90}{\includegraphics[width=1.25\textwidth]{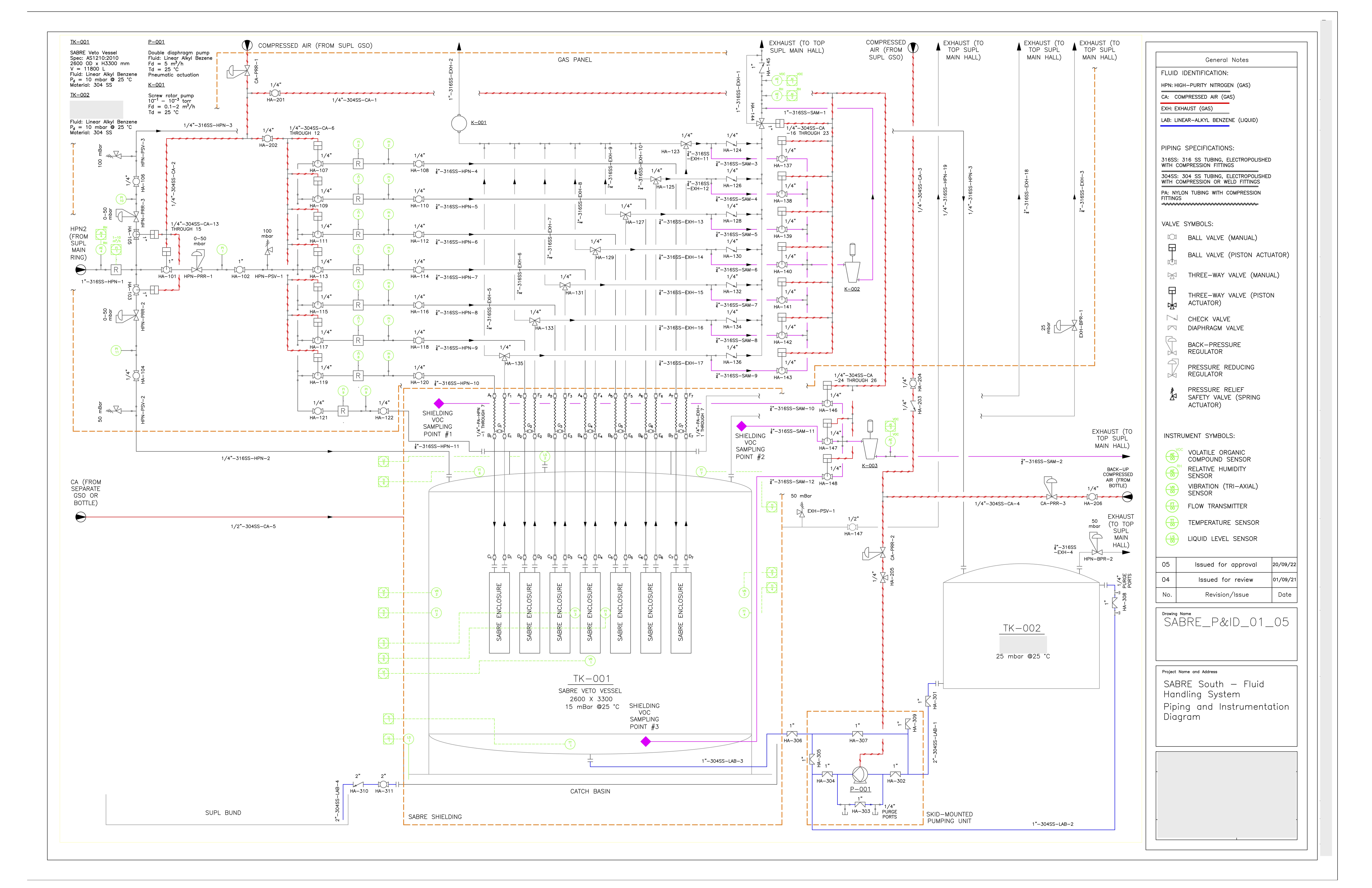}} 

    \caption{Piping and instrumentation diagram of the  SABRE South fluid handling system. Details of the system are given in the text.}
    \label{fig:GHTechnical}
\end{figure}

A gas handling system has been developed to satisfy these requirements throughout the life of the experiment. The gas handling system features three components: a supply panel, an exhaust panel, and an interface panel or staggered gas manifold. The latter connects the supply to the main detector through each subsystem volume, and can be temporarily disconnected to install the crystal insertion system during deployment and extraction of the crystal enclosures. The supply line is connected to the main SUPL supply ring. A technical diagram showing each of the three main panels is shown in Fig.~\ref{fig:GHTechnical}.

\subsubsection*{Exhaust system}
The exhaust panel collects the HPN exhausted from the crystal enclosures and the veto vessel. A set of manual three-way valves redirects the exhaust flow towards either the main exhaust line or the pumping line. During normal operation, the gas is exhausted through the main exhaust line. During initial cleaning of the gas lines and after each crystal insertion operation, the individual crystal enclosure lines can be redirected towards the pumping line, so that vacuum and backfilling cycles can be run through one or more crystal enclosures to restore an appropriate contamination level. The staggered gas manifold features a bypass line for each crystal enclosure, so that the crystal enclosures can be included or excluded from the pumping/backfilling cycles.

The main exhaust section features an automated gas manifold through a compressor. This allows for loop sampling of gas traces (i.e. a sampling procedure in which a sample is removed from the enclosure, tested, and put back into the enclosure) in the crystal enclosures, to ensure the absence of leaks and contaminants that would harm the crystals.
Each individual crystal enclosure exhaust line splits into a fork. The sampled gas can be redirected to the individual lines through a compressor. A set of seven compressed air-actuated ball valves is controlled by a programmable logic controller. Relative humidity and volatile organic compound sensors are integrated into the system.

\begin{figure}[htb]
    \centering
    \includegraphics[width=0.6\textwidth]{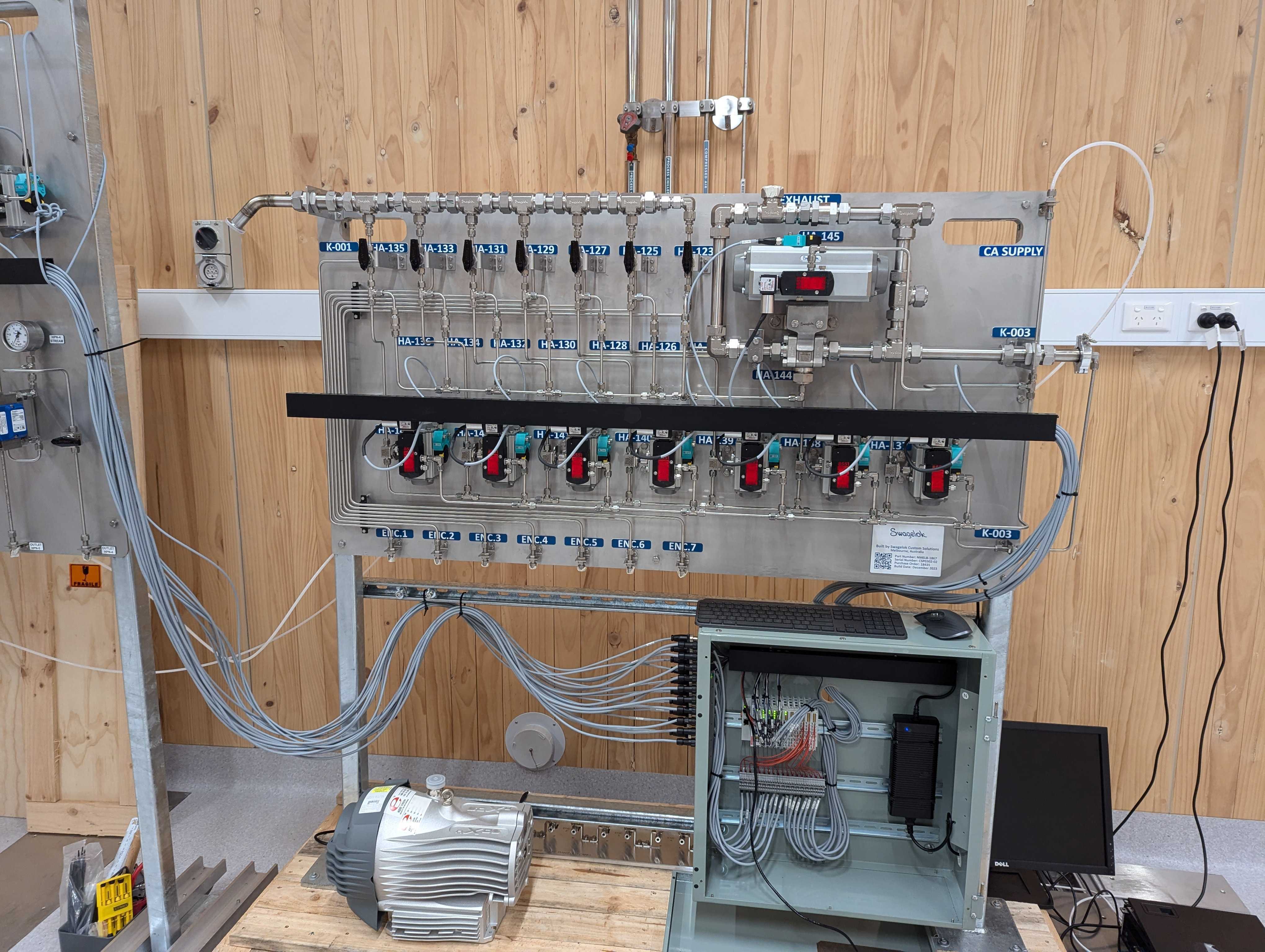}
    \caption{The exhaust system gas panel undergoing laboratory testing at the ANU.}
    \label{fig:GasManifold}
\end{figure}

\subsubsection*{Staggered gas manifold}
The staggered gas manifold is integrated into the gas handling system to allow flexibility during crystal insertion system operations.
Each time a crystal enclosure is inserted into or extracted from the veto vessel, the corresponding HPN supply and exhaust lines must be disconnected to allow the crystal insertion system glove box to be mounted above the top flange of the vessel.

\section{Liquid scintillator veto}
\label{sec:SABRE_veto}

SABRE South's active veto system serves to identify background events originating from radioactive decays within the crystals that may mimic a signal and to act as passive shielding for background external to the crystals. For example, $^{40}$K, a key impurity, decays through the emission of an Auger electron with an energy in the range of 1--6 keV coincident with a 1.5~MeV  $\gamma$ ray. This decay process proceeds via K-shell electron capture, by which the $^{40}$K becomes an excited state of $^{40}$Ar and emits a 3~keV X-ray and an Auger electron, followed by the decay of the excited state and the emission of a 1.5~MeV $\gamma$ ray. Previous experiments have accounted for this by employing double-coincidence rejection, where it was assumed that the escaping gamma ray would be detected by another crystal in the sensitive mass~\cite{Bernabei:2009}. However, this detection mechanism is not fully efficient, as demonstrated by the peak at around 3~keV in the DAMA data where events from $^{40}$K are expected~\cite{Bernabei2018}.

\begin{figure}[ht]
    \centering
    \includegraphics[width=0.7\textwidth]{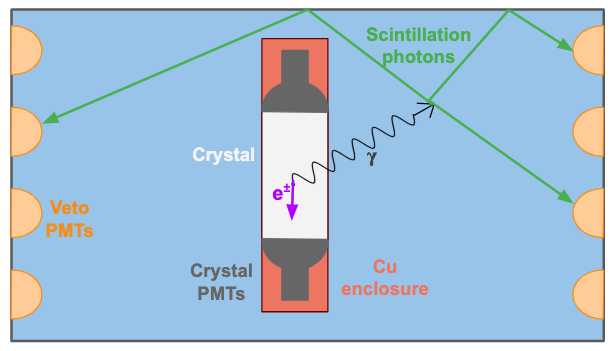}
    \includegraphics[width=0.28\textwidth]{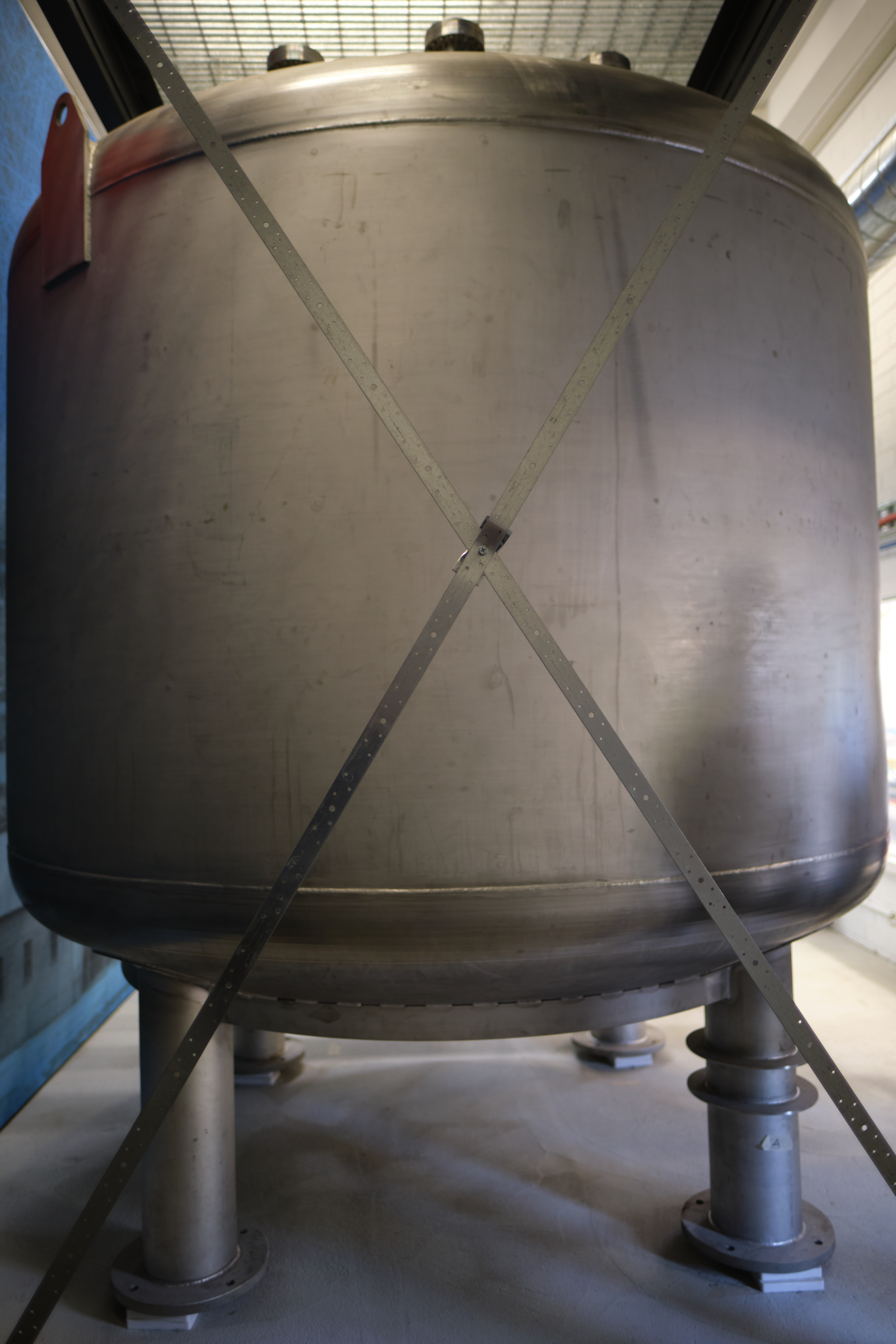}
    \caption{Left: Illustration of the active veto concept. A decay product in the crystal (purple) can be identified by simultaneous detection of other decay products (green) observed by the veto PMTs. Right: An image of the steel veto vessel, in storage before its move to SUPL.}
    \label{fig:active_veto}
\end{figure}

By immersing the NaI(Tl) crystal detector modules in a liquid scintillator (Fig.~\ref{fig:active_veto}), high-energy $\gamma$ rays produced by radioactive decays can be detected and the decay event vetoed. The SABRE South full background simulation demonstrated that the veto can reduce background from important $^{40}$K contamination by a factor of 10 in the region of interest and suppress the total experimental background by 27\%~\cite{sabre_background}. While the $^{210}$Pb contamination has a large contribution to the total background, it is not possible to veto as $\alpha$ particles are predominantly emitted along the decay chain, as well as low energy $\gamma$s that do not reach the liquid scintillator.

The trigger logic for the veto is being designed to optimise veto efficacy with minimal deadtime, with information such as multi-PMT coincidence as well as charge-based and time-based information.

SABRE South uses approximately 11,600 L of liquid scintillator that is a mixture of linear alkylbenzene (LAB), sourced from the supplier to the JUNO experiment, doped with approximately 3.5~g/L of the fluorophore PPO (2,5-diphenyloxazole) and 15~mg/L of bis-MSB (1,4-bis(2-methylstyryl) benzene). The fluorophore concentrations are subject to further optimisation to ensure low absorption~\cite{Anderson_2021}. Purification of the liquid scintillator is not planned during operation. The veto detector uses Hamamatsu 204~mm R5912 PMTs with oil-proof potted bases. The nominal design uses a baseline of 18 PMTs, but it is possible a total of 32 could be installed with the inclusion of PMTs from the decommissioned Daya Bay experiment. The PPO and Bis-MSB are required to mitigate the self-absorption properties of LAB \cite{Anderson_2021} and maximise the total light yield. The wavelength of light output of LAB alone is around 300~nm \cite{lab-lambda}. In the presence of fluorophores the maximum wavelength of emission is 422~nm, which is closer to the peak quantum efficiency of the PMTs. The key components of the liquid scintillator veto detector are summarised in Table~\ref{tab:sabre-specs-veto}.

\begin{table}[htb]
\centering
\caption{Components of SABRE South liquid scintillator veto system with their specifications.}
\label{tab:sabre-specs-veto}
\begin{tabularx}{0.99\textwidth}{ll}
\toprule
\bf{Component}   & \bf{Specifications}                             \\ 
\midrule
Scintillator    & 11,600~L Linear alkylbezene (LAB) \\
Fluorophores      & $\sim$3.5~g/L PPO \& $\sim$15~g/L Bis-MSB\\
Readout         & CAEN V2730 digitiser, 32~channel, 500~MS/s, 0.12 mV resolution\\
Vessel          & Stainless steel, 3~m (height) $\times$ 2.6 m (diameter)       \\ 
Vessel lining   & Lumirror reflector        \\ 
Photomultipliers& Hamamatsu R5912 w. oil-proof base, 204~mm~diameter, \\ 
                & arranged in 3-6-6-3 configuration       \\
\midrule
\end{tabularx}
\end{table}

In addition to reducing the background, the veto system provides additional passive shielding from external radiation, increasing the distance and density of material between the crystals and external components. Based on the simulations of Ref.~\cite{sabre_background}, the background contributions from both the liquid scintillator itself and anything external to it are reduced by more than 99\% due to this detector system. This leads to a contribution of approximately < ${\cal O}(10^{-4})$~cpd/kg/keV$_{\rm{ee}}$ from any component external to the detector and its shielding.

 The liquid scintillator is effective in rejecting events in which electrons, neutrons, and photons deposit more than 20 keV of energy in the scintillator. For the baseline scenario with 18 veto PMTs, this enables the detection of energy depositions with 50\% efficiency at a threshold of 20 keV, rising to 99\% at 80 keV, based on simulations of energy depositions from crystal background sources in the veto.
This expected performance is important if the DAMA/LIBRA signal is to be excluded, as it provides a means of understanding what background processes might cause a DAMA/LIBRA-like modulation through the use of techniques such as pulse-shape discrimination. This has been tested conceptually using a liquid scintillator prototype detector, containing approximately 40~L of LAB-based liquid scintillator and a single R5912 PMT, in which neutron and $\gamma$ ray signals could be discriminated. This suggests that the same approach could be applied to the full liquid scintillator veto, particularly since the arrival of 16 additional PMTs from the decommissioned Daya Bay experiment gives us the opportunity to increase the number of PMTs in the detector from 18 to 32, thereby increasing photosensor coverage and light yield/light collection efficiency. Given a 0.75~PE/keV position averaged detection probability for detection by any given PMT, an energy threshold for pulse shape discrimination can be estimated to be as low as 150--200~keV. This depends on the method of pulse shape discrimination, with advanced multivariate machine learning techniques likely to reduce the threshold. Position reconstruction in the liquid scintillator veto is also a goal for the experiment and is being explored. This is useful for identifying events in the veto that may have originated from the crystals, thereby aiding the veto process via fiducialisation of a volume around the crystals.

\subsection{Vessel design}

\begin{figure}[htb]
    \centering
    \includegraphics[width=0.48\textwidth]{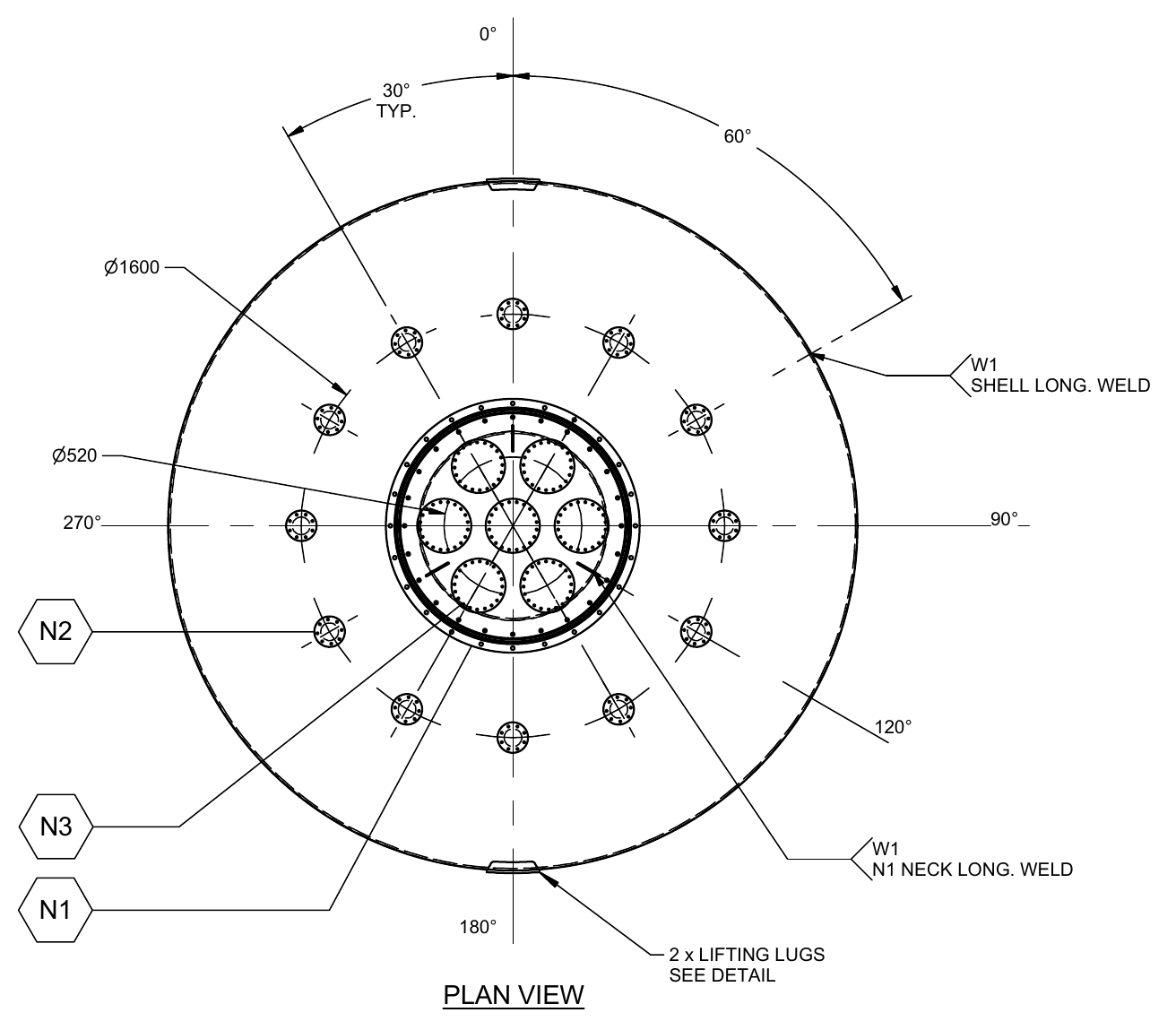}
    \includegraphics[width=0.8\textwidth]{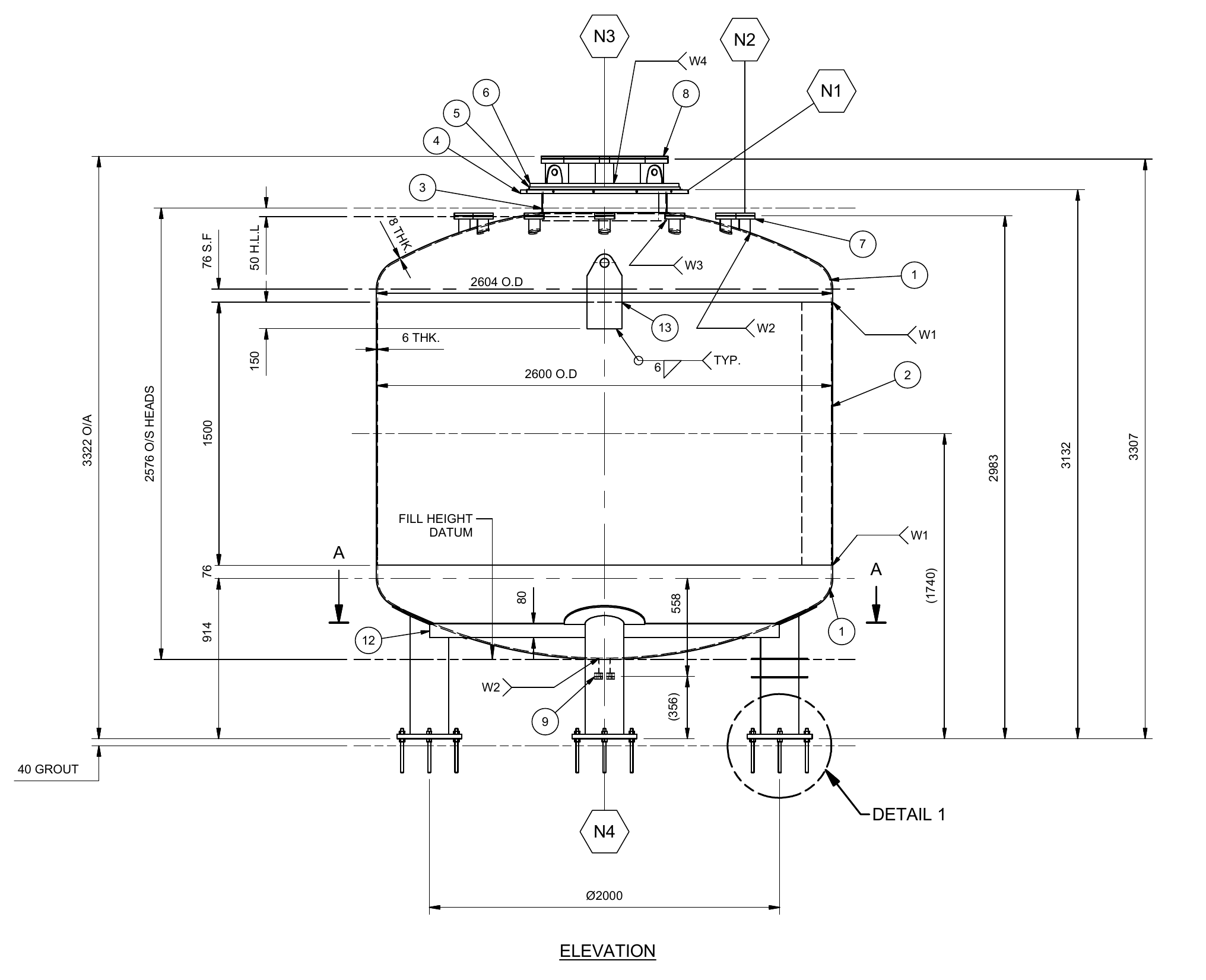}
    \caption{Top down and side view of the veto vessel.}
    \label{fig:vessel}
\end{figure}

\begin{figure}[tbh]
\centering
\includegraphics[width=0.6\textwidth]{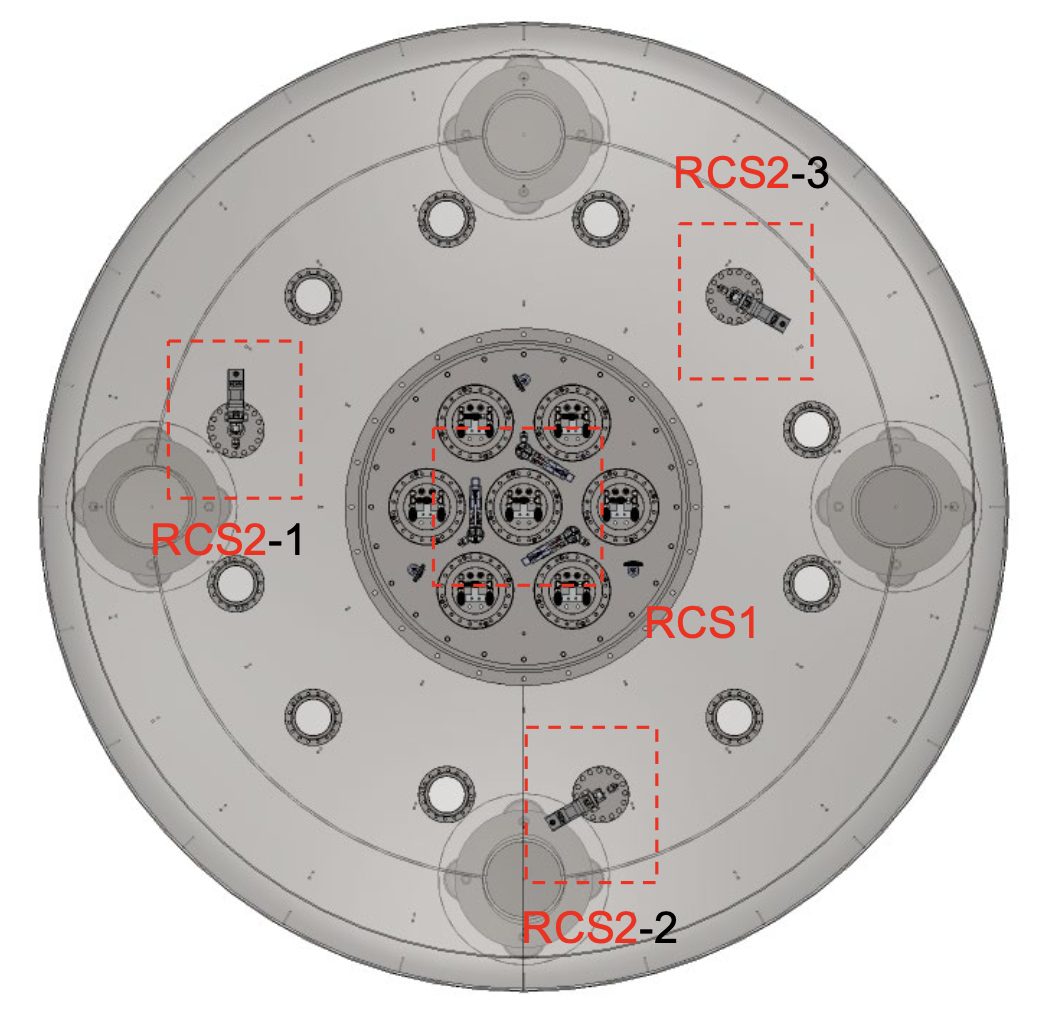}
  \caption{Top view of the veto vessel, also showing the top flange at the centre of the render. The positions of the radioisotope calibration systems of crystal detectors (RCS1) and veto sytem (RCS2) are highlighted in red.}
  \label{fig:RCSpos}
\end{figure}

The vessel structure can be divided into three sections: the central barrel and the two torispherical caps located at the top and bottom of the vessel. To ensure that the vessel is small enough to be safely transported to SUPL on the back of a vehicle and large enough for MeV-scale photons to be fully absorbed inside the liquid scintillator, an outer diameter of 2600~mm (at the middle of the barrel) and a total height of 3322~mm (inclusive of the top flange) were chosen. A diagram of the vessel is shown in Fig.~\ref{fig:vessel}. Rows of stud bolts are welded to the internal wall of the vessel to mount the PMTs and reflector. A central flange (``top flange'', see Fig.~\ref{fig:RCSpos}), is located at the top of the vessel, which is flanked by a circle of smaller side flanges. The central flanges are used for the insertion of the crystal enclosures, fluid handling, cable feedthroughs, and calibration systems. 

The top flange of the vessel serves as the point of insertion of the crystal detector system. The crystal enclosures are inserted via the seven minor flanges arrayed around the top flange, the crystal detector radioactive calibration system is inserted via the three smaller flanges surrounding the central crystal module, and related cables are connected via feedthrough tubes attached to the flanges. Figure~\ref{fig:RCSpos} shows a top down view of the vessel, with the top flange at the centre of the vessel. The top flange is made of stainless steel, with an outer diameter of 850~mm, and a thickness of 20~mm. The seven minor flanges are identical in dimensions and design, with an outer diameter of 202~mm and arrayed with a collection of mounting holes for the cable feedthroughs. The calibration pipe flanges have an outer diameter of 36~mm. 

The SABRE South veto vessel was designed in-house and manufactured and certified by Tasweld Engineering in 2018. Medical-grade steel was used for the fabrication of the vessel. The steel was screened with HPGe detectors at LNGS: the mean activity of key background radionuclides are reported in Table~\ref{tab:bulk_steel}. All welds were made with non-thoriated welding rods to prevent the deposition of additional radioactive material.

\begin{table}[htb]
\centering
\caption{Average radioactivity levels measured on samples of the stainless steel used for the SABRE veto vessel. For $^{40}$K, $^{137}$Cs and $^{235}$U, no contamination was found and the averages of the $90\%$ confidence level upper limits are provided instead. The measurements were performed at LNGS with HPGe detectors.\label{tab:bulk_steel}}
\begin{tabular}{cc}
\toprule
Isotope & Activity [mBq/kg] \\
\midrule
$^{238}$U & 1.5\\
$^{232}$Th & 2.0\\
$^{235}$U & $< 1$\\
$^{40}$K & $< 8$\\
$^{60}$Co & 5\\
$^{137}$Cs & $< 0.5$\\
\bottomrule
\end{tabular}
\end{table}

\subsection{Assembly}

There are three main components installed inside the vessel: the reflective Lumirror lining,  the PMTs on their mounts,  and the optical calibration system. They are installed in the following stages: (i) installation of the Lumirror, and its securing/bracing to the vessel wall; (ii) installation of components of the calibration system; (iii) pre-assembly of PMTs on to their mounts; (iv) installation of PMT mounts on the vessel walls.

\begin{figure}[htb]
    \centering
    \includegraphics[width=0.47\textwidth]{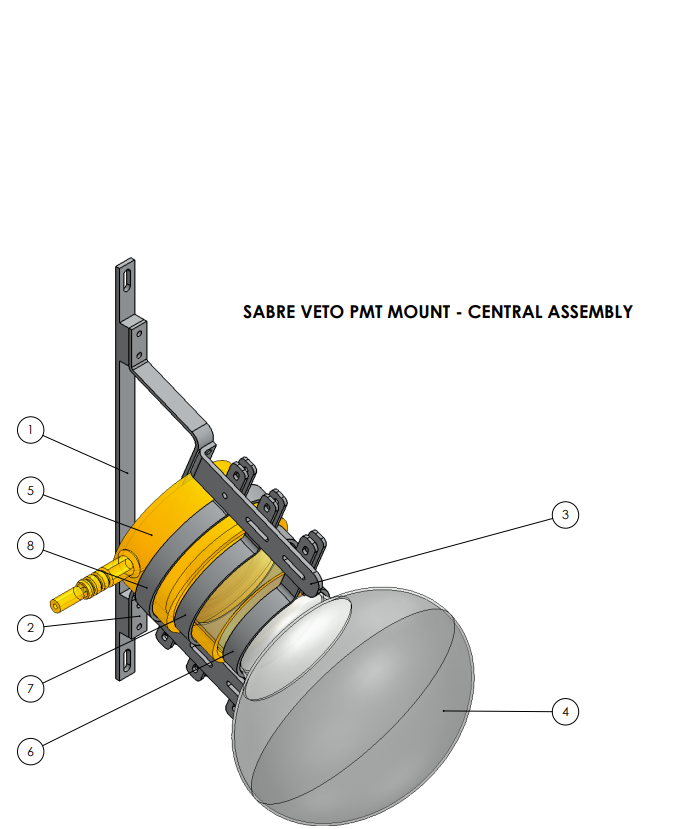}
    \includegraphics[width=0.47\textwidth]{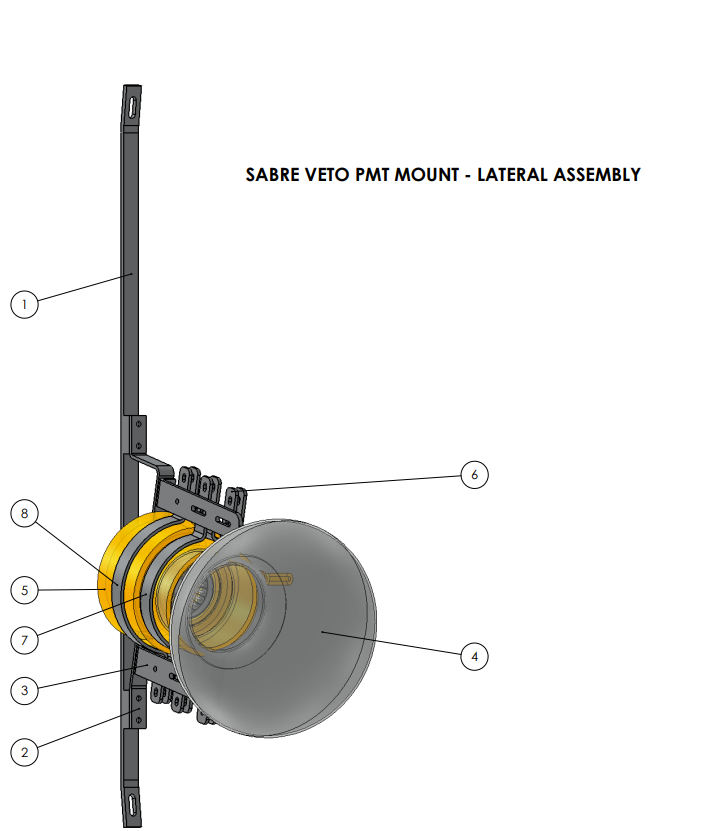}
    \caption{Engineering models of the central and lateral mounting units. The mounts are nominally made of aluminium. Labels 1--3 highlight the central mount frame; labels 4 and 5 point out the bulb and oilproof base of the R5912 respectively; and labels 6--8 indicate the clamps in which the PMT is mounted. The mounts will be fastened to two stud bolts, at the top and bottom of each mounting bracket, whilst the clamps will be fastened with stainless steel nuts and bolts.}
    \label{fig:PMT_Mounts}
\end{figure}

The Lumirror sheets are cut to size based on pre-assembly measurements. The PMT mounts are designed so that they can be fastened to the vessel stud bolts and will clamp onto the PMT in three different positions: two on the oil-proof base and one on the neck of the PMT. The technical diagrams of the mounts are shown in Fig.~\ref{fig:PMT_Mounts}, of which there are two models. The lateral assembly is fastened to the torispherical top and bottom caps, and the central assembly is fastened to the central barrel.

\subsection{Cleanliness protocols}

The vessel must be pickled internally because an oxide layer is expected to have formed between manufacturing and assembly, which may contain radioactivity and may not be compatible with the liquid scintillator. Pickling is performed using a corrosive solution with a concentration that is balanced enough to corrode the oxide layer and not the steel itself. Following this, the vessel is rinsed multiple times with a solution of de-ionised water and Alconox detergent, followed by a rinse with de-ionised water. Rinsing ceases once the outgoing water/solution reaches a required resistivity of $\geq 14$~MOhm~cm. The R5912 veto PMTs and Lumirror, along with all other internal components, follow a similar cleaning procedure to the R11065 PMTs, which is also performed in a cleanroom environment. The installation of internal components is performed in a clean room-like environment with radon reduction, with all installation personnel wearing clean-suits, boot-covers, gloves, and hairnets. Pumping and transfer of the liquid scintillator into the vessel will be performed clean, with the liquid scintillator sparged with HPN once the vessel has been filled. The liquid scintillator was transported and stored in a new ISO tank made of 316L steel, and blanketed in nitrogen. The transferral and cleaning of this tank was handled by the liquid scintillator producer, which was direct from the distillation port and acceptably clean. In case of contamination of the liquid scintillator during filling or transfer we will pause the filling process. There is a total of 17~kL of liquid scintillator in the ISO tank, and since only 11.6~kL is required for the LS veto. The remainder can be used for liquid transfer commissioning.

\subsection{R5912 PMTs}
Owing to the sparse array of PMTs relative to the volume of liquid scintillator, the probability of photoelectron (PE) detection is 0.2--0.3 PE/keV/PMT. Consequently, the pre-calibration of these PMTs is driven by the need to be sensitive to low numbers of PEs, to inform our simulation/digitisation and thus analysis, and also by the ability to deploy a degree of position reconstruction and particle ID in the liquid scintillator veto. With such reconstruction capabilities, we can disentangle our background contributions and pursue other physics using the veto sub-detector --- for example, studies of neutron spallation induced by muon tracks, or potential detection of supernova neutrinos (in which tagging the neutron of an inverse beta decay could improve our detection efficiency). There is no magnetic shielding surrounding these PMTs.

Signal-related characteristics measured for the R5912 PMTs are: single-photoelectron (SPE) charge and gain, single-photon detection efficiency (quantum efficiency), timing properties, and determination of saturation onset. The noise characteristics of the PMTs are being studied, including dark rate over a range of temperatures, afterpulse rates, and spontaneous light emission from the oil-proof bases of these PMTs (behaviour that has been previously reported in Ref.~\cite{DoubleChooz:2016ibm}).
   
\section{Muon veto}
The cosmogenic muon flux modulates with the seasons due to changes in temperature and pressure in the atmosphere. High-energy muons, with energies above 250 GeV, can penetrate rock and reach the experiment. These muons can interact with the detector material, producing spallation neutrons that can mimic the dark matter signal modulation.  SABRE South uses an array of solid plastic scintillators on top of the shielding (Fig.~\ref{fig:sabre-schem}) as both a muon veto system and to measure the muon flux throughout the year. While it is independently triggered, data from the muon system can be combined with the LS veto system muon flux measurements. 

\begin{table}[htb]
\centering
\caption{Components of the SABRE South muon system with their specifications, where TTS denotes the transit time spread.}
\label{tab:sabre-specs-muon}
\begin{tabularx}{0.99\textwidth}{lX}
\toprule
\bf{Component}                                     & \bf{Specifications}                             \\ \midrule
Detectors & 8 panels, 300 cm $\times$ 40 cm $\times$ 5 cm, with fishtail light guides \\
Scintillator & EJ-200 \\
Photomultipliers & Hamamatsu R13089, 2 per module, 46~mm dia. photocathode, 230 ps TTS\\
Readout & CAEN V1743 digitiser, 3.2$\times10^9$ samples/s\\
\midrule
\end{tabularx}
\end{table}

\begin{figure}[htb]
    \centering
    \includegraphics[width=0.6\textwidth]{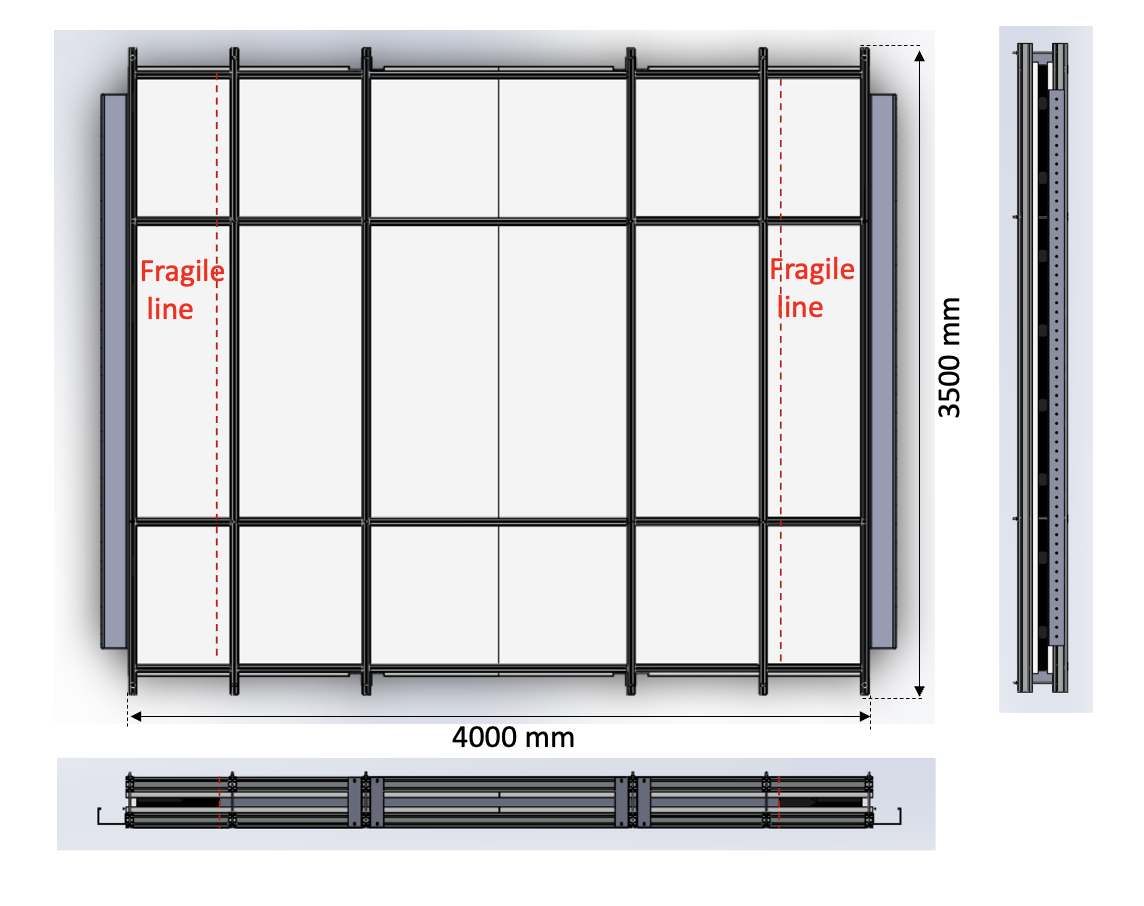}
    \includegraphics[width=0.34\textwidth]{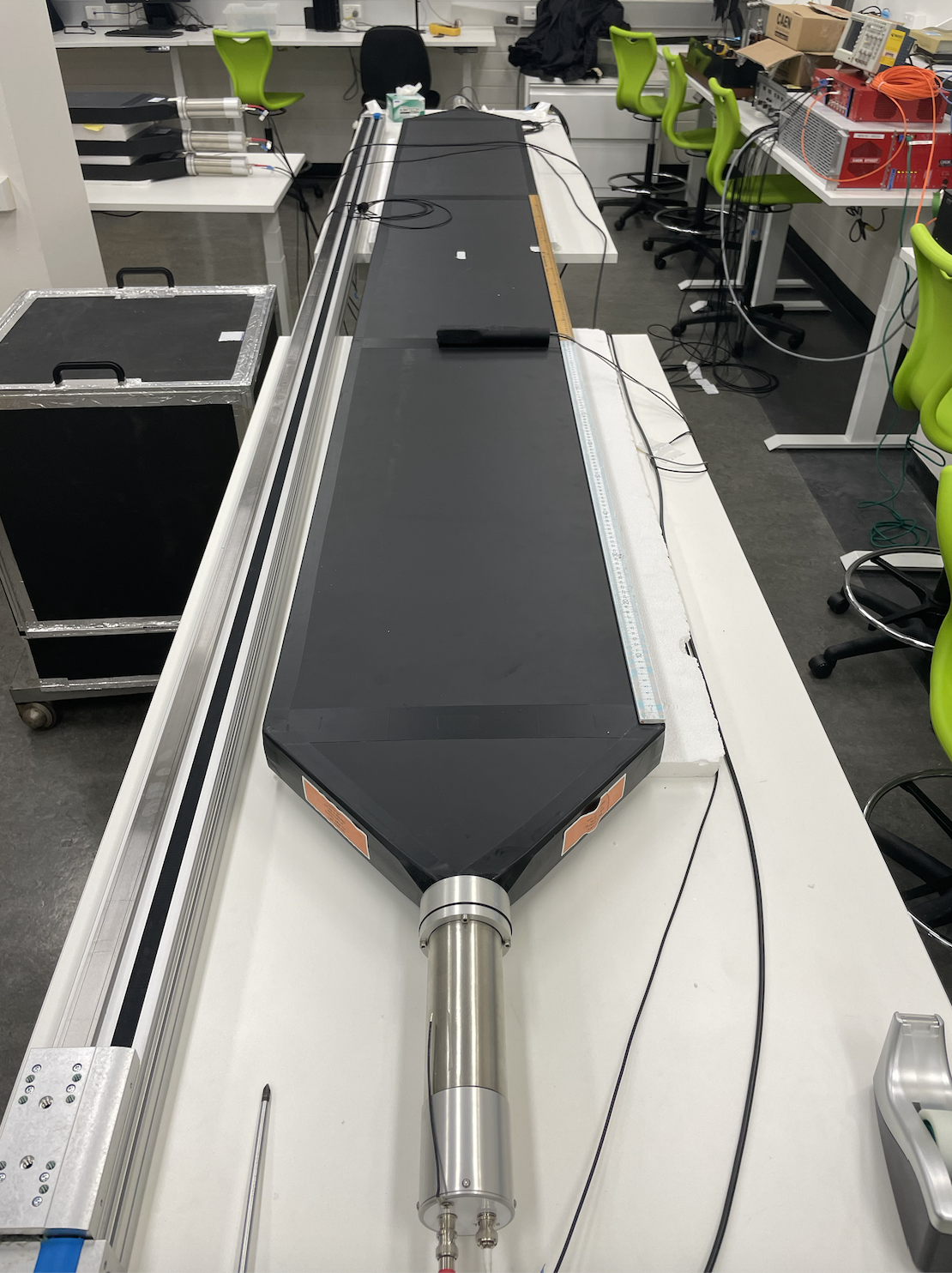}
    \caption{Left: A schematic of the muon veto support frame itself, with fragile lines marked. Right: Image of a single muon detector, with the linear calibration stage placed to the left.}
    \label{fig:muondet}
\end{figure}

\begin{figure}[htb]
    \centering
    \includegraphics[width=0.6\textwidth]{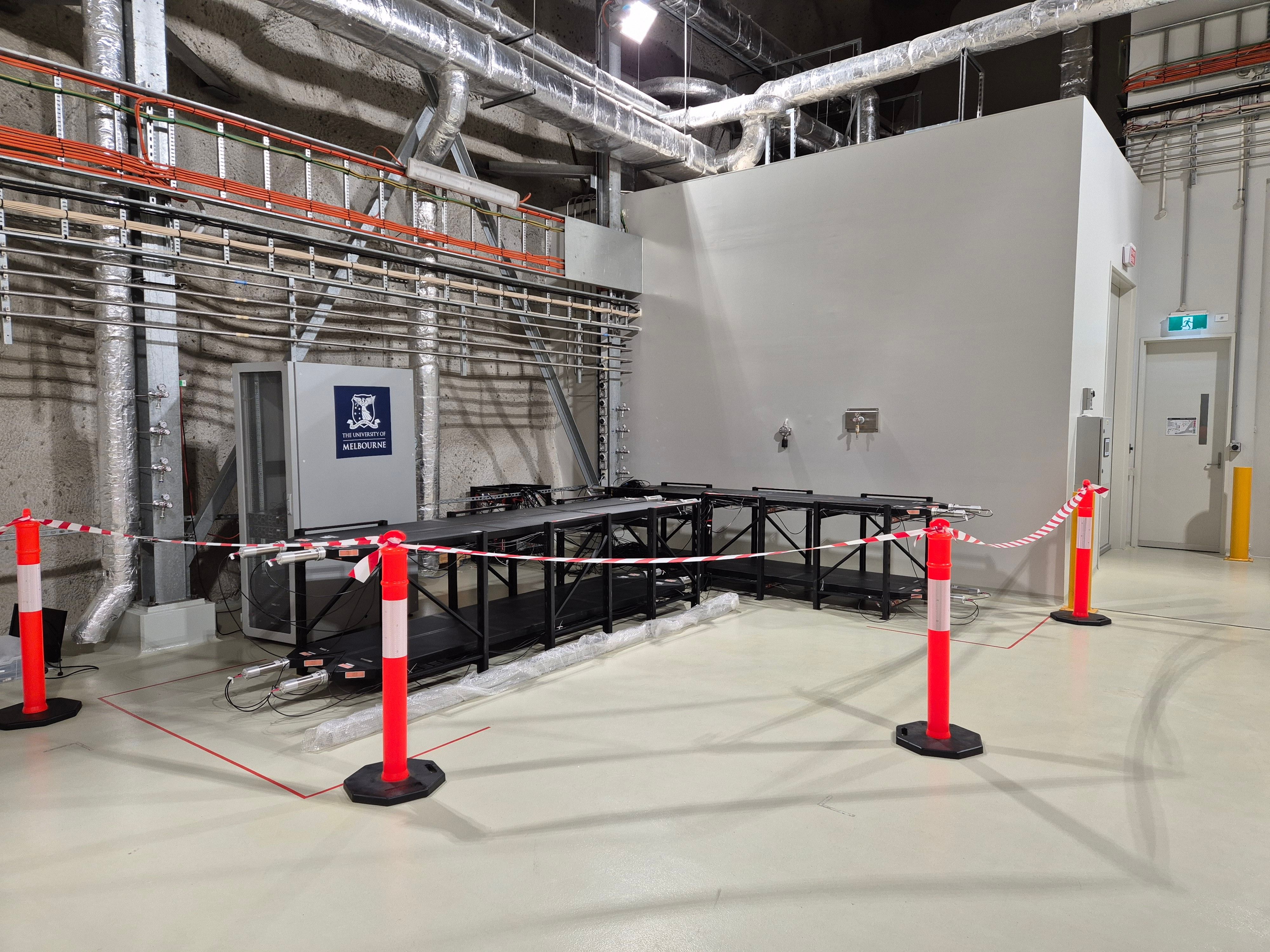}
    \caption{Muon detectors in SUPL, assembled in `telescope' mode, which is the configuration required to measure the muon flux and angular spectrum (from two directions).}
    \label{fig:muonsSUPL}
\end{figure}

The 9.6~m$^2$ (3~m~$\times$~3.2~m) muon detector system is made of eight modules of EJ-200 solid plastic scintillator from ELJEN Technology~\cite{EJ200}, with CAEN Electronics components for readout and the high-voltage system (see Table~\ref{tab:sabre-specs-muon}). 
The sensitive material for each muon detector is a 3000~mm $\times$ 400~mm $\times$ 50 mm EJ-200 plastic scintillator (appropriate given the attenuation length of the EJ-200 is 380~cm). Scintillation photons are transmitted through the trapezoidal PMMA light guides and collected by two Hamamatsu R13089 PMTs, one on each end as shown in Fig.~\ref{fig:muondet}. To protect the joints between the scintillator and light guide, an aluminium frame, as shown in Fig.~\ref{fig:muondet}, will be used. The red dashed lines show where these fragile joints are located. Polystyrene or EPMD rubber will be placed between the muon detectors and the aluminium profiles to keep the detectors in place. HV and signal cables for the muon system will be secured and connected to a patch panel on the frame. Crane lugs on the frame will allow for the entire muon system to be removed as a single entity between the roof of the radiation shielding, and a parking frame within reach of the DAQ rack.
On top of the detector sits a 3.4~m motorised linear stage to control a source-based calibration system. The system has been fully commissioned and calibrated at the University of Melbourne. The muon system is operating in SUPL to start measurements of the cosmogenic background and its modulation throughout the year, Fig.~\ref{fig:muonsSUPL} shows the muon detectors underground in SUPL. 
A rendering of these detectors from above is shown in Fig.~\ref{fig:muondet}.

\begin{figure}[htb]
    \centering
    \includegraphics[width=0.44\textwidth]{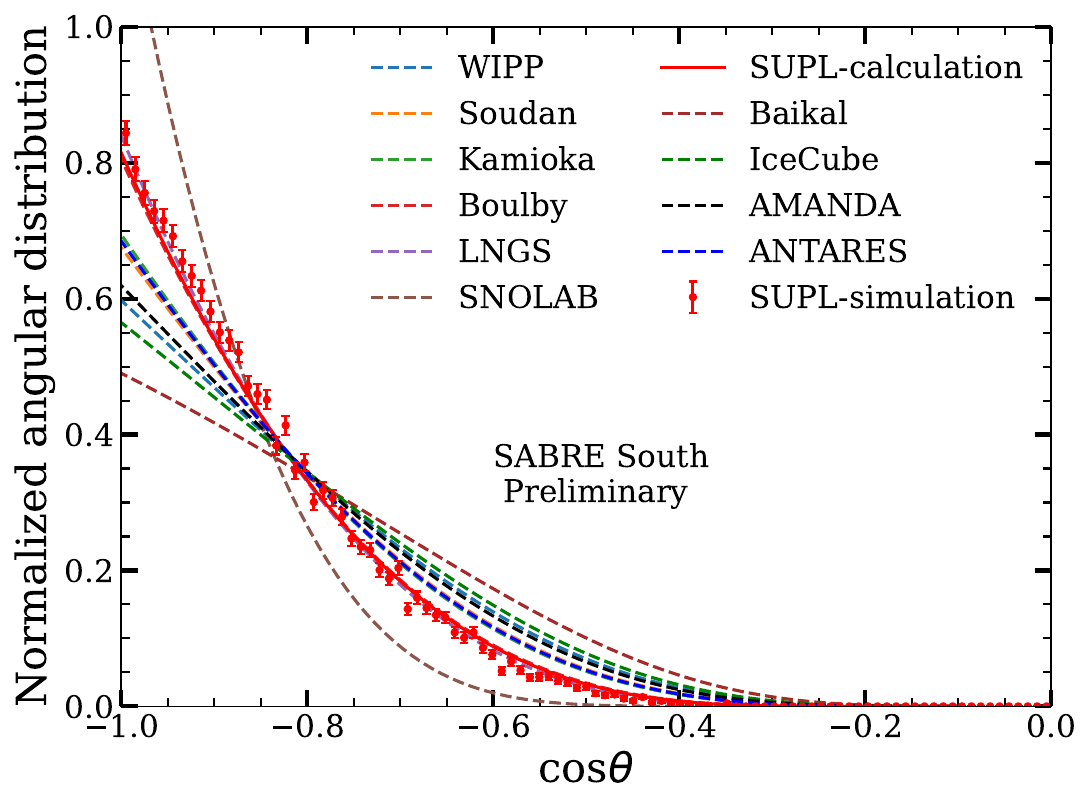}
    \includegraphics[width=0.54\textwidth]{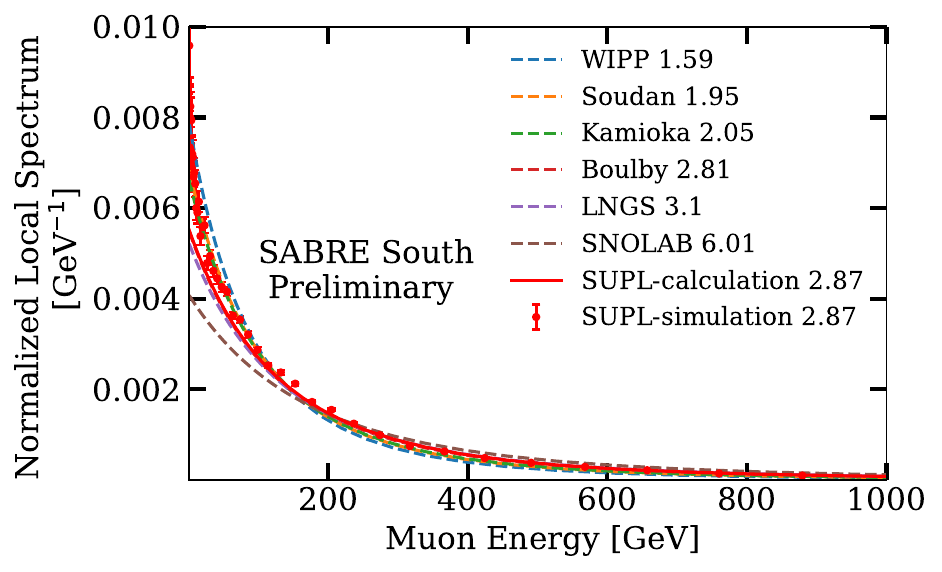}
    \caption{Left: Simulation results showing the normalised angular spectrum of muons that reach SUPL, which we aim to experimentally verify with the muon detectors assembled in `telescope' mode in SUPL. One could use angle projected by the size of the muon veto and this distribution to estimate the number of muons that will hit the liquid scintillator, and miss the muon detectors. These muons will be vetoed by the liquid scintillator instead. Right: The simulated energy distribution of muons that reach SUPL. This can be expected to be the energy range of muons to be tagged by the muon veto, although the muon detectors will not be used as a spectrometer.}
    \label{fig:MuonDists}
\end{figure}

The digitiser used for the readout is a CAEN V1743, which was specifically chosen for its high sample rate of 3.2$\times10^9$ samples/s (which complements the TTS of the PMTs) and resolution of $<$ 8~ps RMS~\cite{CAEN_1743}. The timing resolution of this system allows for accurate position reconstruction along the length of each muon paddle, where the time difference between signals in each PMT is used to determine position. Measurements performed on the surface have found that this can be performed at a position resolution of approximately 5~cm. The efficiencies of each detector have also been measured on the surface to be at least 99\% for muons that fully traverse the thickness of the detector. The surface flux and statistical uncertainty at the University of Melbourne was measured to be 131.3 $\pm$ 0.4~s$^{-1}$m$^{-2}$. At SUPL we expect the muon rate for each panel (1.2~m$^2$), with statistical uncertainties, to be 37.8 $\pm$ 4.3~d$^{-1}$, and 302.7 $\pm$ 34.0~d$^{-1}$ for all eight panels with an overall coverage of 9.6~m$^2$, based on a measurement with a smaller detector. Figure~\ref{fig:MuonDists} shows the expected energy and angular distributions for muons reaching SUPL compared to other underground labs, where the distributions are based on calculations from models provided in Ref.~\cite{MuonCalcBkg}, and the points for SUPL are derived from simulations using Geant4~\cite{geant4Updates, GEANT4} and CRY~\cite{CRY}. Muons will always be vetoed, and there should be no muon background in the crystals. Highly angular muons may miss the muon veto, but will still be vetoed by the liquid scintillator. Likewise, if vertical muons miss the liquid scintillator by passing through conduits/other detector material, the vertical coverage provided by the muon detectors will provide the veto.

The muon detector uses Hamamatsu R13089 PMTs, which have a peak quantum efficiency near a wavelength of 420 nm, which matches the EJ-200's maximum emission wavelength of 425~nm.

Gain stability studies have been conducted in multi-day runs on the surface. The PMTs can take one to two days to stabilise, during which the gain can vary by approximately 5\%. The PMTs are gain matched in each detector module and monitored using radioactive source-based calibration and muon signals.

   \section{Calibration systems}

The SABRE South sub-detectors require periodic calibration, which is particularly important for accurate time-dependent measurements. Calibration refers to the process of bias correction and verification of the accuracy of a detector's measurements. The liquid scintillator veto and NaI(Tl) calibration systems for SABRE South are designed for remote operation, and are located within the shielding structure.
The response of the detector to particle interactions can change with time, due to: (i) environmental factors such as temperature, humidity, mechanical stress, and vibrations (monitored by the slow control system outlined in Sec.~\ref{sec:slowcontrol}); (ii) electrical changes, including variations in the supply voltage and electromagnetic interference; (iii) detector ageing, where over time the optical properties of scintillators can deteriorate (e.g. reduced absorption length or light yield), and the PMTs can experience changes in gain, detection efficiency and other features used for pulse shape discrimination.

SABRE South uses radioactive sources to calibrate the NaI(Tl) detectors, the liquid scintillator veto, and the muon detector. Additionally, the liquid scintillator veto is calibrated using an optical (pulsed laser light) system. The two calibration systems for the liquid scintillator veto are complementary: the radioactive calibration system provides accurate high-energy calibration lines, while the optical system is capable of sending low-intensity, high-frequency pulses to test the single photoelectron and timing response of the PMTs. Calibration with radioactive sources requires data runs of a few hours; other NaI(Tl) experiments typically have 3–4 hour calibration runs every 2 weeks~\cite{Amare2019eff}. The laser calibration can be conducted more frequently in short bursts, as it does not require any mechanical adjustments to the system. The frequency of calibration for both systems will be adjusted as needed to optimise detector stability and live-time.

\subsection{Radioisotope calibration system}

To calibrate the detectors, radioactive sources must be placed close enough to produce a signal distinguishable from other detected signals during calibration runs. However, during normal data acquisition, the signal from the calibration source must be kept negligible. This is managed by keeping the radioactive sources shielded and at a distance from the detectors during normal operations, then moving them closer to the detectors only during calibration runs. Simulations show that by using sources with sufficiently low activity and ensuring adequate shielding, the dead time of the veto is minimally affected. This approach eliminates the need to disassemble the shielding for each calibration run, maximising detector live-time while maintaining efficient calibration.

\begin{figure}[tbh]
\centering
\includegraphics[width=0.41\textwidth]{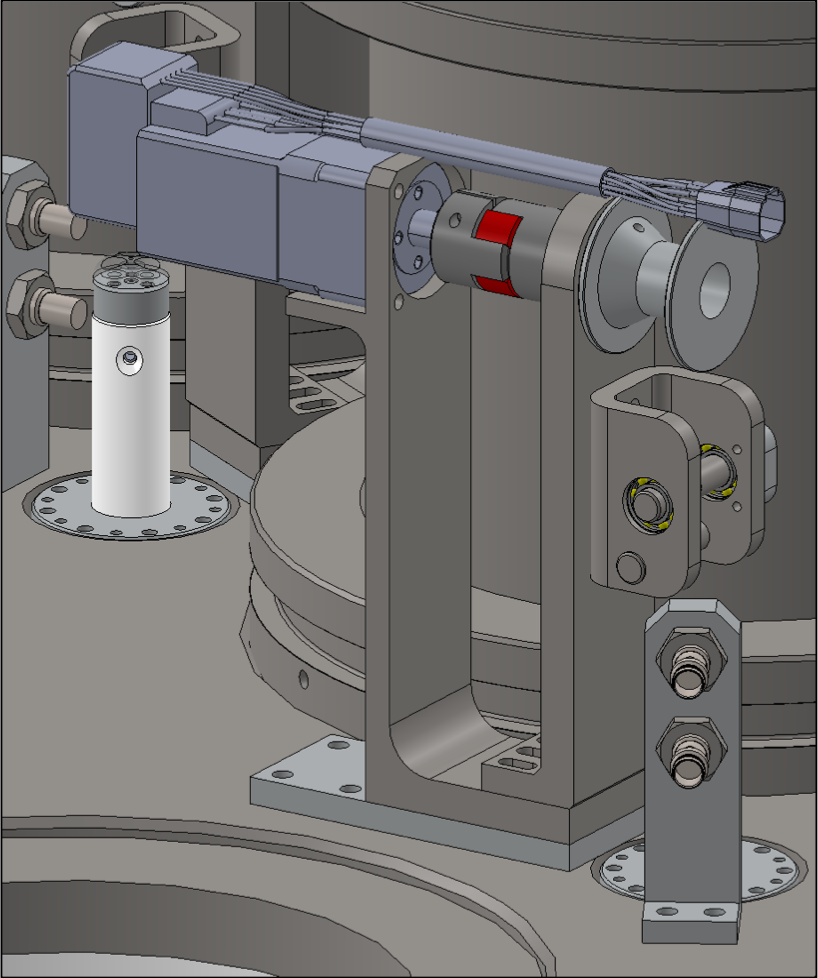}
\includegraphics[width=0.445\textwidth]{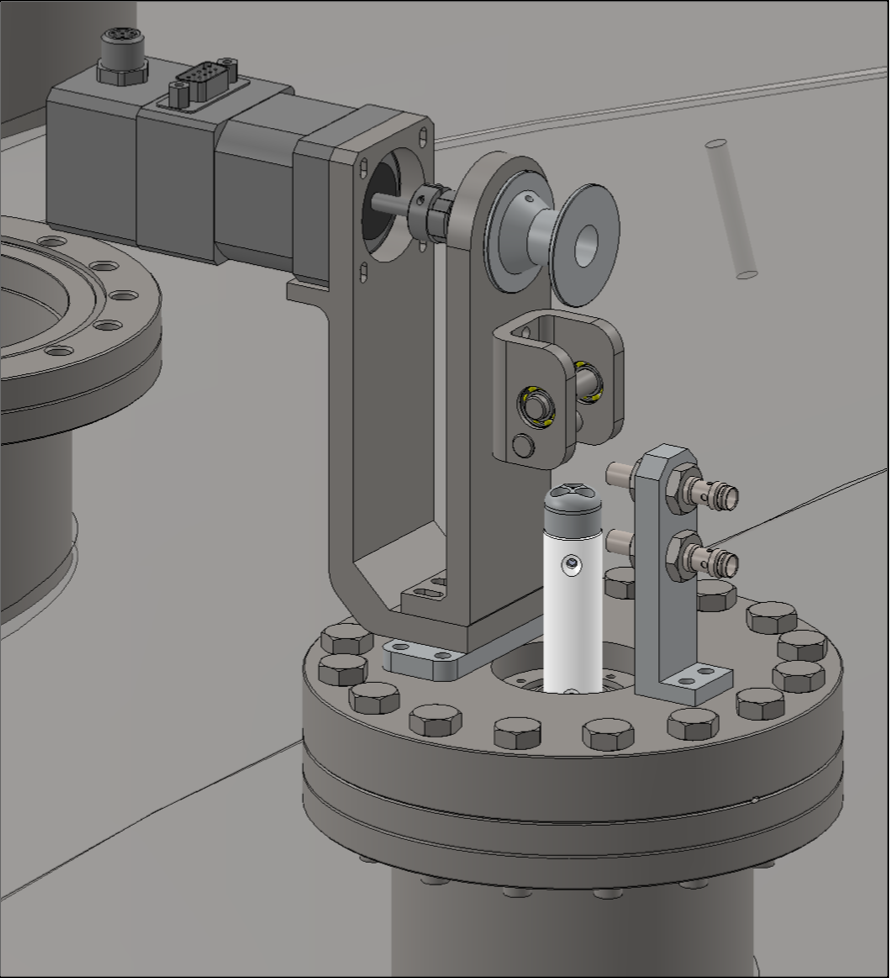}
  \caption{Close-up views of the radioisotope calibration systems for (left) the crystal detectors and (right) the liquid scintillator veto system.}
  \label{fig:RCS}
\end{figure}

SABRE South employs motorised calibration deployment systems for the precise calibration of both the crystal detectors and the liquid scintillator detectors, referred to as RCS1 and RCS2, respectively (Fig.~\ref{fig:RCS}). 

Each radioactive calibration system consists of a steel mount, a stepper motor, a calibration probe suspended by a wire, an alignment reel, and position measurement sensors.  The stepper motor controls the winding and unwinding of the wire around a spool, enabling the precise vertical movement of the calibration probe. The wire is guided through a centring reel, ensuring the probe slides accurately into an aluminium pipe that directs it to the calibration position.

The systems are operated via the slow control system using the EtherCAT protocol, which allows radioactive sources to move between the calibration and normal operation positions in minutes. This process is fully automated. Access to the system will only be necessary in rare cases of equipment failure.

Three units of each motorised calibration deployment system are installed on top of the veto vessel, as shown in Fig.~\ref{fig:RCSpos}, so that the detectors are evenly irradiated and the PMT signals are comparable. Each triplet of RCS units is equidistant from the axis of the vessel and 120 degrees apart. The RCS1 units are positioned so that each calibration source is equidistant from a triplet of crystal detectors when lowered into the calibration pipe (150~mm from the crystal axis).

The RCS2 units are mounted on top of the three outer flanges (151.6~mm diameter) at a distance of 800~mm from the centre of the vessel. Under consideration is the installation of a steel cylinder (151.6~mm external diameter, 16.8~mm internal diameter and 100~mm height) on top of the flange to house the calibration sources and shield their radiation during normal data acquisition. During the calibration runs, the sources of each RCS2 system are lowered to the vessel mid-plane (the same height as the centre of the crystals) at a position equidistant from pairs of veto PMTs. Each veto PMT is at most 1.5~m from any radiation source. The RCS2 calibration probes are designed to enter the vessel through aluminium pipes, ensuring that the LAB remains sealed and isolated from both the calibration source and the external atmosphere.

\begin{figure}[htb]
    \centering
    \includegraphics[width=0.485\textwidth]{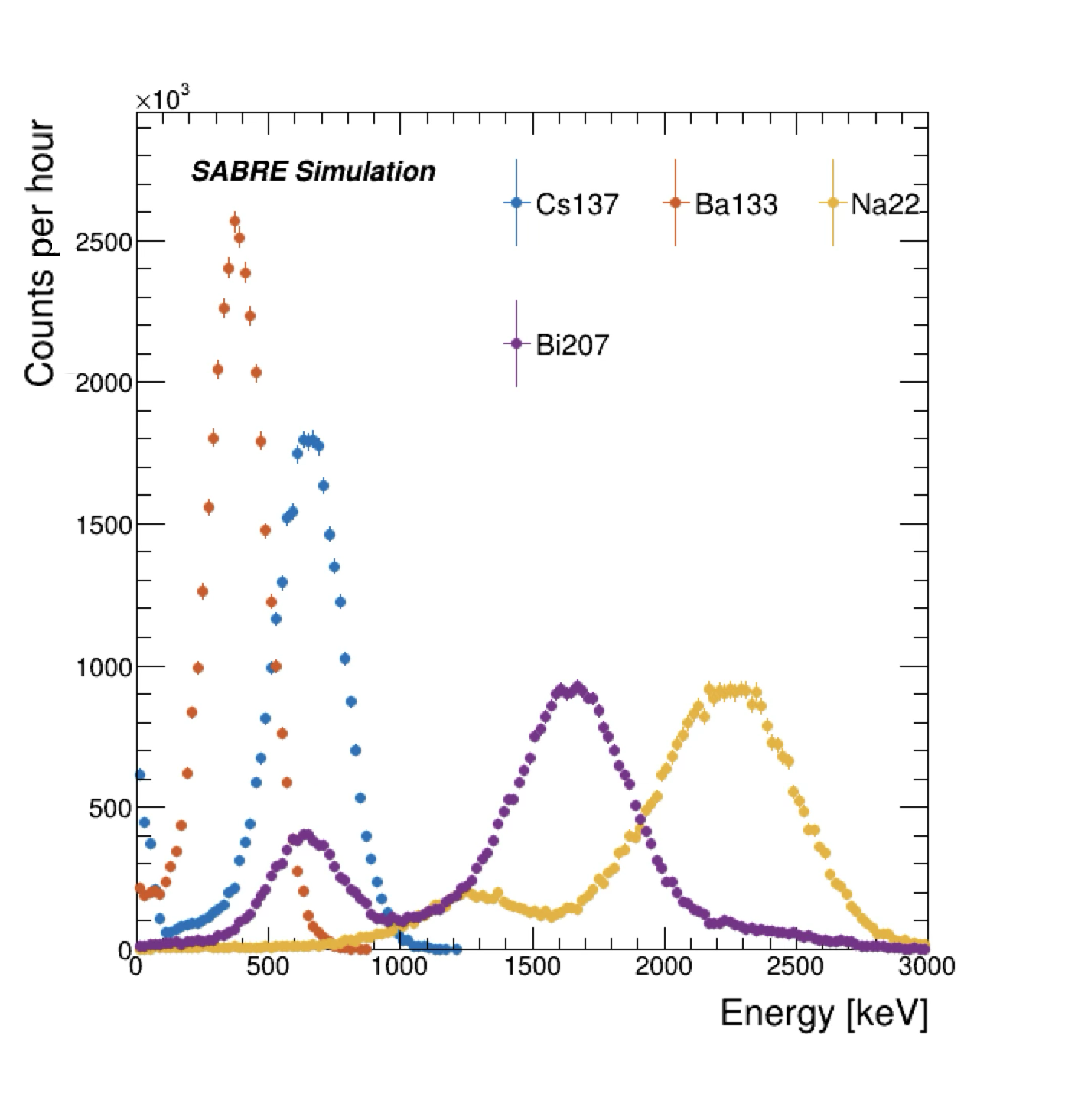}
    \includegraphics[width=0.485\textwidth]{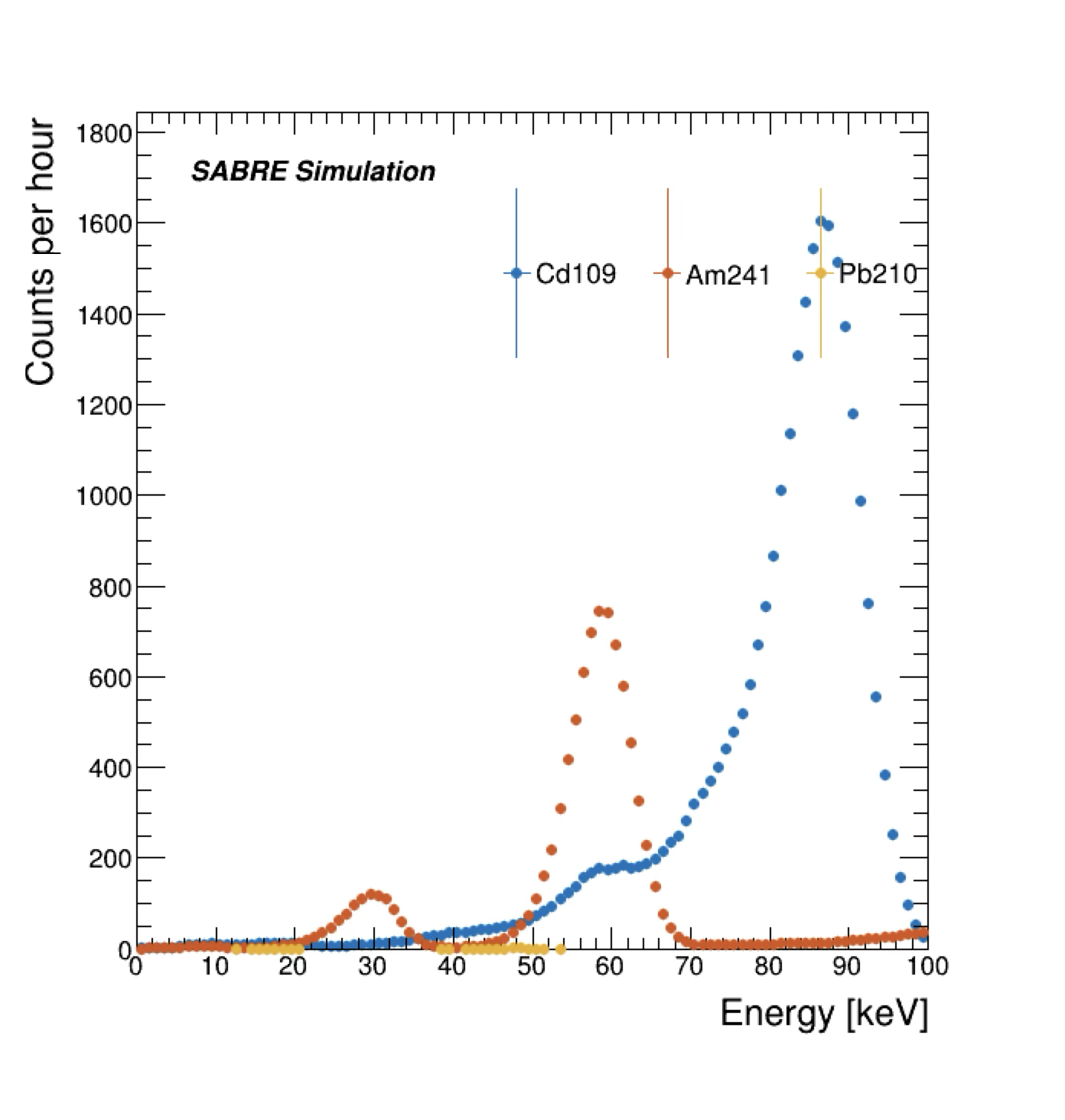}
    \caption{The left figure shows the energy depositions for the sources used to calibrate the LS detector, with the detector resolution taken into account. The right figure shows the energy depositions for the sources used to calibrate the crystal detectors, also considering the detector resolution. Both sets of spectra have been smeared to match the expected resolution of each detector and are zero-suppressed, with error bars representing the statistical uncertainty. The distributions are normalized to reflect the number of disintegrations expected in one hour for 10 kBq sources. The inclusion of $^{210}$Pb was a test to evaluate its viability as a potential source.}
    \label{fig:RCSSources}
\end{figure}

The RCS1 system uses the following radioactive sources $^{109}$Cd and $^{241}$Am, each with an activity of approximately 10 kBq. This activity level is chosen because the background radiation is small compared to other sources of background. For RCS2, the system uses higher-energy sources, $^{137}$Cs, $^{133}$Ba, $^{22}$Na, and $^{207}$Bi. These sources produce more significant background signals, so their activity is kept below 1 kBq to minimise this effect. The simulated energy depositions for each system are shown in Fig.~\ref{fig:RCSSources}, which illustrates the energy depositions for the RCS1 system using a single crystal. The muon veto is calibrated with a $^{60}$Co source. 

\subsection{Optical calibration system}

The veto detector system is monitored and calibrated using optical light from a pulsed laser system. This calibration involves assessing the PMT timing, gain, resolution, optical properties of the LAB scintillator, and overall detector stability. The conceptual design of the optical calibration system utilises optical fibres, which are sheathed in Teflon tubes, to transport picosecond laser pulses to specific emission positions within the veto vessel.

Two diffusive bulbs (or laser balls) are attached to steel calibration tubes, which are then connected to ConFlat 151.6~mm outer diameter flanges on the veto vessel. Quartz optical fibres are chosen to carry signals from the laser light source, as the material is chemically compatible with LAB. The optical fibres are sheathed with PTFE to increase reflectivity and improve signal isotropy.
The laser balls are nominally positioned at two different vertical and horizontal positions within the vessel. This configuration allows for studies of single photoelectron signals with each PMT. They are made of PTFE with radii of 25~mm and are a proven solution for liquid scintillator experiments~\cite{JunoOptCalib}. The fibres are attached to the laser balls using connectors made of LAB-compatible polyether-ether-ketone resin (PEEK)~\cite{JunoOptCalib}. Studies on shadowing effects caused by the connector have shown that this effect is negligible. The stainless steel tubes provide stability to the system.

The laser is located outside the detector, with an optical splitter dividing the beam into three fibres: two fibres are inserted into the vessel, and one connects to a reference sensor system. The nominal light source is an LDB-200 picosecond pulsed laser from Tama Electric, which can be remotely controlled. A wavelength of 405 nm was initially considered due to the high QE of the PMTs at this wavelength. However, the photon absorption length at this wavelength is 0.3~m. As a result, a laser wavelength of 445 nm, where the absorption length is 26~m, is chosen. The PMT QE is only 5\% lower than at 405~nm.

The optical calibration system is simulated with Geant4 to ensure that it is suitable for purpose. The simulation has been used to determine the number of photoelectrons collected by each PMT, and to examine the expected photon flight times, trajectories in the LAB, and their probability of survival.

   \section{Data Acquisition and slow control}

\begin{figure}[htb]
    \centering
    \includegraphics[width=\columnwidth]{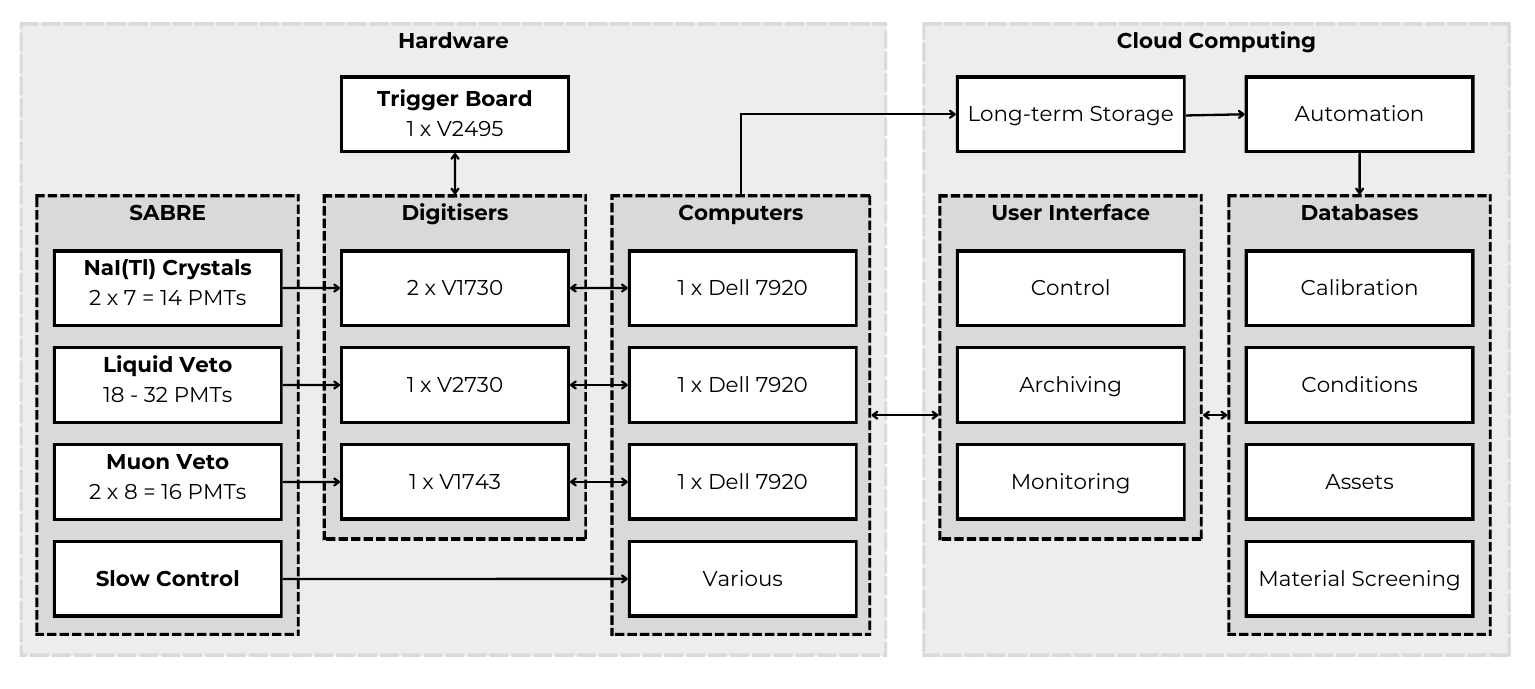}
    \caption{Schematic layout of the SABRE DAQ, monitoring, and control systems. Note that this is for the nominal crystal readout not including preamplifiers. Dual-readout with preamplifiers on the crystals would require 2 additional V1730 digitisers.}
    \label{fig:DAQ-SABRESystems}
\end{figure}

\begin{figure}[htb]
    \centering
    \includegraphics[width=0.33\columnwidth]{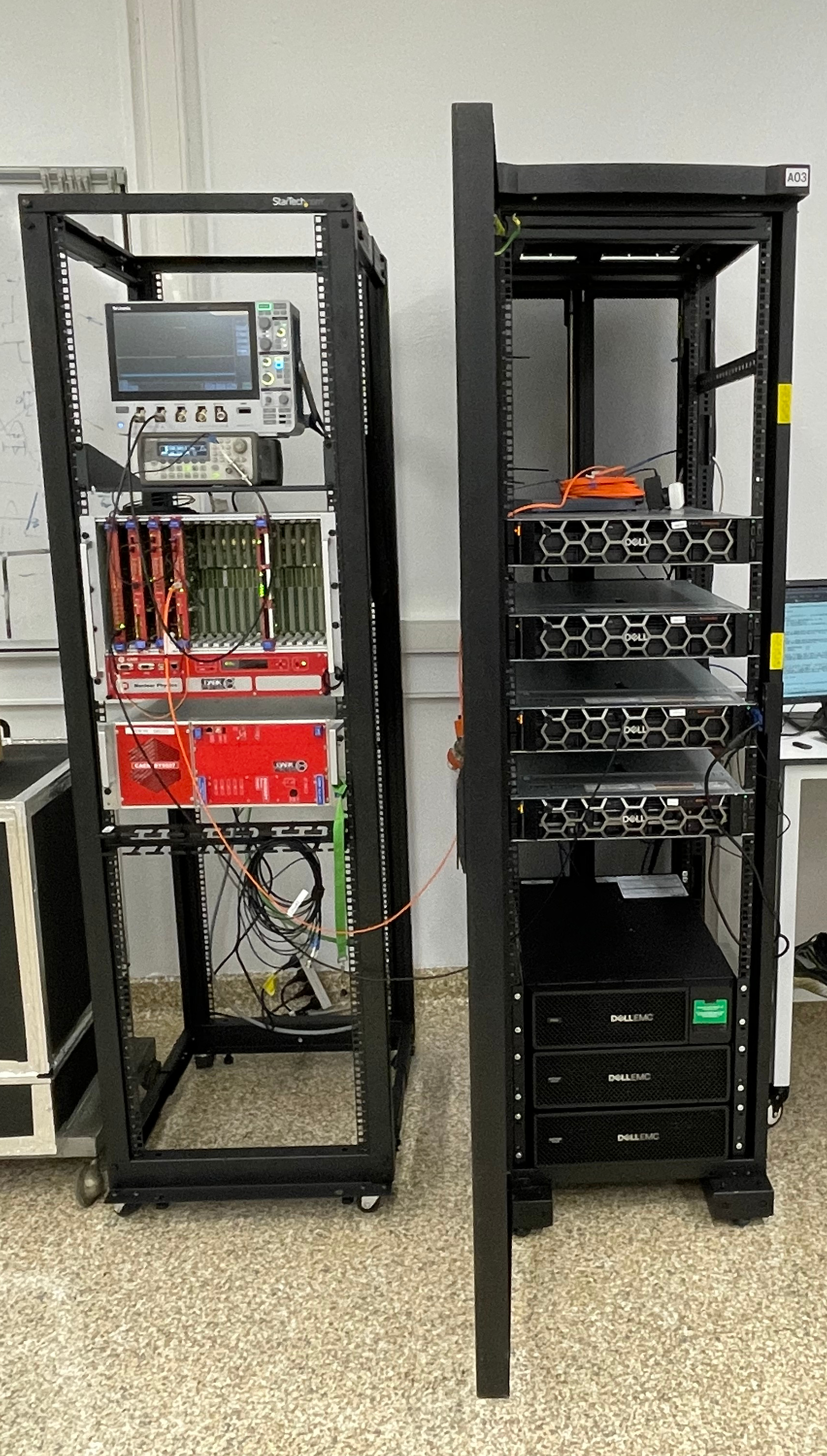}
    \caption{SABRE DAQ electronics (left) and server units (right).}
    \label{fig:DAQ-DAQRack}
\end{figure}

A schematic layout of the Data Acquisition (DAQ) system is given in Fig.~\ref{fig:DAQ-SABRESystems} as well as other related components that are discussed in more detail in later sections. Signals from every PMT are digitised by CAEN digitisers, and the resulting data are then transferred to a computer and written to disk. Each of the SABRE subsystems (the crystal, liquid scintillator, and muon detectors) is managed largely independently with three DAQ instances collecting data in parallel. For the crystal subsystem, a dual-readout approach is being considered in which two signals from each PMT are digitised (amplified and unamplified). A CAEN FPGA (Field Programmable Gate Array)-based board, running custom firmware, is used to coordinate triggering and provide synchronisation signals between the digitisers where required. A custom C++ program is responsible for managing the data collection from the digitisers. 
The system described here is operational in SUPL and used in pre-calibration (Fig.~\ref{fig:DAQ-DAQRack}).

Live data-taking systems are managed through custom run-control software built on React Automation Studio (RAS). This uses the networked publisher-subscriber interface provided by EPICS (Experimental Physics and Industrial Control System) to control and coordinate the various systems. The slow control interfaces through a dedicated extensible interface to EPICS for control and a database system for data archiving. Grafana~\cite{Web:Grafana:Docs} and Prometheus~\cite{Prometheus} are used for in-depth monitoring.

\subsection{Digitisers}
\label{sec:DAQ-Digitizers}

Several CAEN digitisers are used to record the signals generated by the detector PMTs: the 14 crystal PMTs are digitised by two V1730 digitisers (8 channels per digitiser); the liquid scintillator PMTs use a V2730 digitiser (32 channels); and the 16 muon PMTs use a V1743  digitiser (16 channels per digitiser). Table~\ref{tab:digitisers} summarises the digitisers used for each sub-detector.

\begin{table}
  \centering
  \caption{Summary of digitisers used by each sub-detector.} 
  \label{tab:digitisers}
\begin{tabularx}{0.99\textwidth}{lXXXX}
    \toprule
    & \bf{Digitiser}                                     & \bf{Channels} & \bf{Sampling rate (S/s)} & \bf{Resolution (bits)} \\ \midrule
    \bf{Liquid scintillator veto}            & V2730 & 32    & $500 \times 10^6$ & 14 \\
    \bf{Crystal detectors}  & V1730 & 8     & $500 \times 10^6$ & 14\\
    \bf{Muon veto}          & V1743 & 16    & $3.2 \times 10^9$ & 12 \\
    \midrule
\end{tabularx}
\end{table}

\subsubsection*{V1730}
\label{sec:DAQ-V1730}

The CAEN V1730~\cite{CAEN_1730} is an 8-channel, 500~MS/s digitiser with a 14-bit ADC and a 0.5 or 2~Vpp (peak-to-peak) dynamic range. It is capable of continuous acquisition with zero dead time.  Whenever the digitiser is triggered, the time of the event is recorded along with waveforms for each channel and some meta-data.
To achieve an energy threshold of 1 keV, the trigger must be sensitive to single photoelectron signals. The noise rate at such a threshold demands the use of coincidence triggers across pairs of PMTs coupled to the same crystal and configured as channel pairs on the digitiser.

\subsubsection*{V2730}

Used to record signals and waveforms generated in the liquid scintillator veto, the 32-channel V7230 digitiser samples at 500~MS/s with a 14-bit ADC and a dynamic range of 4~Vpp~\cite{CAEN_2730}. The V2730 also has improved triggering options through custom FPGA firmware that this new generation of CAEN digitisers supports, allowing individual channel triggers, and correlated triggers between channels.

\subsubsection*{V1743}

The CAEN V1743~\cite{CAEN_1743} is a 16-channel, 3.2~GS/s digitiser with a 12-bit ADC and a 2.5~Vpp dynamic range.  It is used with the muon detector sub-system where the high-digitisation rate (and subsequent timing resolution) is useful for position reconstruction. Due to its high digitisation rate, the board has a significant amount of dead time (125~$\mu$s for 1024 samples) per event. Since the expected event rate is $\sim$10 Hz (estimated based on measurements, inclusive of noise, of the muon detectors installed in SUPL) the dead time is not expected to have a significant impact on efficiency.

\subsection{FPGA logic unit}
\label{sec:DAQ-PLU}

To coordinate triggering between the independent digitisers, a CAEN V2495~\cite{CAEN_2495} programmable logic unit (PLU) with a customisable FPGA is used. The trigger conditions of interest to SABRE South include a multiplicity trigger (generated when multiple channels trigger in coincidence) and a singles pre-scale calibration trigger (where only a defined fraction of the single channel events triggered on will be saved).  Additionally, the trigger system is used to provide a common start signal to each of the digitisers, synchronising the start of the acquisition. Custom trigger firmware has been developed.

\subsection{High voltage}

The high voltage to the PMTs is supplied using a CAEN SY5527 \cite{CAEN_5527} Universal Multichannel Power Supply System with A7435 \cite{CAEN_7435} power supplies which can supply up to 3.5 kV and 3.5 mA per channel. Two 24-channel, positive polarity power supplies will be used for the crystal detectors and liquid scintillator veto, whilst a single 24-channel negative polarity power supply will be used for the muon veto. These power supplies have a typical voltage ripple of < 7~mVpp, and voltage set resolution of 5~mV~\cite{CAEN_7435}. The mainframe has a built-in CPU controller to control and monitor each channel. The SY5527 mainframe has a built-in EPICS input-output controller (IOC).

\subsection{Acquisition software}
\label{sec:DAQ-Software}

A custom program has been written to control the SABRE DAQ instances.  Each DAQ system is controlled by independent instances of this program running on the corresponding DAQ computer.  Control and monitoring of these instances is performed through EPICS.

The control flow of the program is as follows. When a start signal is received, the program reads the configuration parameters and initialises each of the boards in the DAQ.  It then enters a loop where each board is checked and any data present are saved directly to file (alternate output streams are also supported).  This continues until either a stop signal is received from EPICS, a pre-set end condition is met (e.g. acquisition duration), or an error occurs.  In addition to the raw data from the digitisers, the acquisition is monitored through a number of EPICS process variables, including event rates, data rates, and sample waveforms.

\subsection{Slow control}
\label{sec:slowcontrol}

Environmental sensors are used not only to monitor the health and performance of the experiment but also to ensure the health of experimenters and detector operators. 
An early prototype read-out system was developed using an NI-cRIO framework for this system \cite{Krishnan2021}. Due to costs associated with expanding the system, an alternative EPICS-integrated and Ethercat-based system has been developed. The new alternative system uses Beckhoff couplers to digitise the analogue signals from each sensor; these are then read using the open-source EPICS Ethercat support module \cite{EPICS_Ethercat}.
The sensors that SABRE South will use are listed below.

\begin{itemize}
    \item \textbf{Thermometers:} Some of the instrumentation is sensitive to temperature, particularly the noise rate of the PMTs. Thus, the temperatures of and within are monitored for later analysis. Resistance thermometer detector (RTD) sensors with four-wire PT100 (platinum 100 ohm) probes are used to measure the ambient laboratory temperature and the temperature of various parts of the SABRE South hardware. These sensors must be far enough from the crystals not to introduce excess background.
    \item \textbf{Barometers:} Air pressure within SUPL is monitored with a Delta Ohm-HD9408.3B piezoresistive sensor, with a resolution of 0.1 hPa. Pressure sensors are important for nitrogen-flushed systems (such as the crystal glove box) to ensure that an adequate flow and overpressure of nitrogen is maintained.
    \item \textbf{Hygrometers:} Relative humidity in the laboratory is monitored using an APAR-AR250 transducer, with a $\pm2$~\% accuracy. A relative humidity sensor is incorporated into the high-purity nitrogen lines that are used to flush the interior of the crystal enclosures and the vessel, to ensure that water is not introduced into the system.
    \item \textbf{Liquid level sensors:} Liquid level sensors are used in two locations: (i) near the top of the inside of the veto vessel to monitor the level of the liquid scintillator; and (ii) in a liquid catch tray beneath the vessel, to directly monitor for any possible leak of LAB from the vessel.
    \item \textbf{Volatile organic compound (VOC) sensors:} A VOC sensor is located within the shielding volume, such that any liquid LAB or LAB vapour leaking from the vessel will be immediately detected. An additional VOC sensor is part of the fluid-handling system to ensure that there are no LAB leaks in the crystal enclosures.
    \item \textbf{Vibration accelerometers:} As SUPL is located in an active gold mine, there are occasions when vibrations from blasts or heavy vehicles elsewhere in the mine may be transmitted through to SUPL, inducing noise in the PMTs. Data on such vibrations are obtained via VSA001 single-axis accelerometers. Three of these are be mounted independently on the outside of the veto vessel, with each sensor oriented to provide x/y/z measurements. Analogue current output (0--10 mA) is proportional to motion (sensitive from 1~Hz to 6~kHz vibration frequency).
    \item \textbf{Radon:} The level of radon in the SUPL environment will be monitored. During experiment operation, we are primarily concerned with radon in the vessel-shielding volume. This will be done with an AlphaGuard DF2000 detector with a measurement range of 2--2,000,000 Bq/m$^3$ and sensitivity of 5 counts per minute at 100 Bq/m$^3$.
    \item \textbf{Oxygen:} Oxygen sensors are placed within the shielding (external to the veto vessel) and within the nitrogen blanket inside the veto vessel (above the LAB level). There may also be oxygen monitoring within the fluid handling system and general SUPL environment. This is necessary as oxygen can dissolve into the liquid scintillator and consequently reduce the light yield.
\end{itemize}

The temperature is controlled via a whole-laboratory solution. The laboratory is cooled to approximately 23~$^{\circ}$C. Humidity and radon within and around the detector elements is controlled via constant flushing of HPN. HPN is constantly flushed through the crystal modules, the volume between the shielding and the vessel, and through a blanket on top of the liquid scintillator within the vessel.

\section{Shielding}
The key elements of the environmental background radiation that must be suppressed in the SUPL laboratory space are radioactive radon gas, neutrons (expected flux of $3-4 \times 10^{-5}$~cm$^{-2}$s$^{-1}$) and gamma rays (expected flux of 2.5~cm$^{-2}$s$^{-1}$) from radioactive decays in the rock and secondary products from cosmic-ray muon interactions in the rock. The liquid scintillator protects the crystals from external radiation, however additional passive shielding is required to: (i) bring the induced background to a negligible level in the crystals; (ii) keep the event rate in the veto low enough to ensure low dead-time; and to allow the use of the veto system for background characterisation and other physics studies such as detection of supernova neutrinos (for a supernova early warning system). 

The SABRE South detector has 112 tonnes of high purity steel and high-density polyethylene (HDPE) shielding. This is made of a 10~cm-thick layer of HDPE, to shield against neutrons, interposed between two 8~cm layers of steel to shield against high-energy gamma rays. This structure is a cube of approximately 4~m side-lengths, and surrounds the main SABRE South liquid scintillator vessel. This configuration was chosen over alternatives, such as stacked water containers or lead, as steel offers excellent structural strength, serves as an effective radiation shield, and is more compact. High purity steel made with over 80\% new steel can also be easily sourced in Australia.

\begin{figure}[tbh]
\centering
\includegraphics[width=0.4\textwidth]{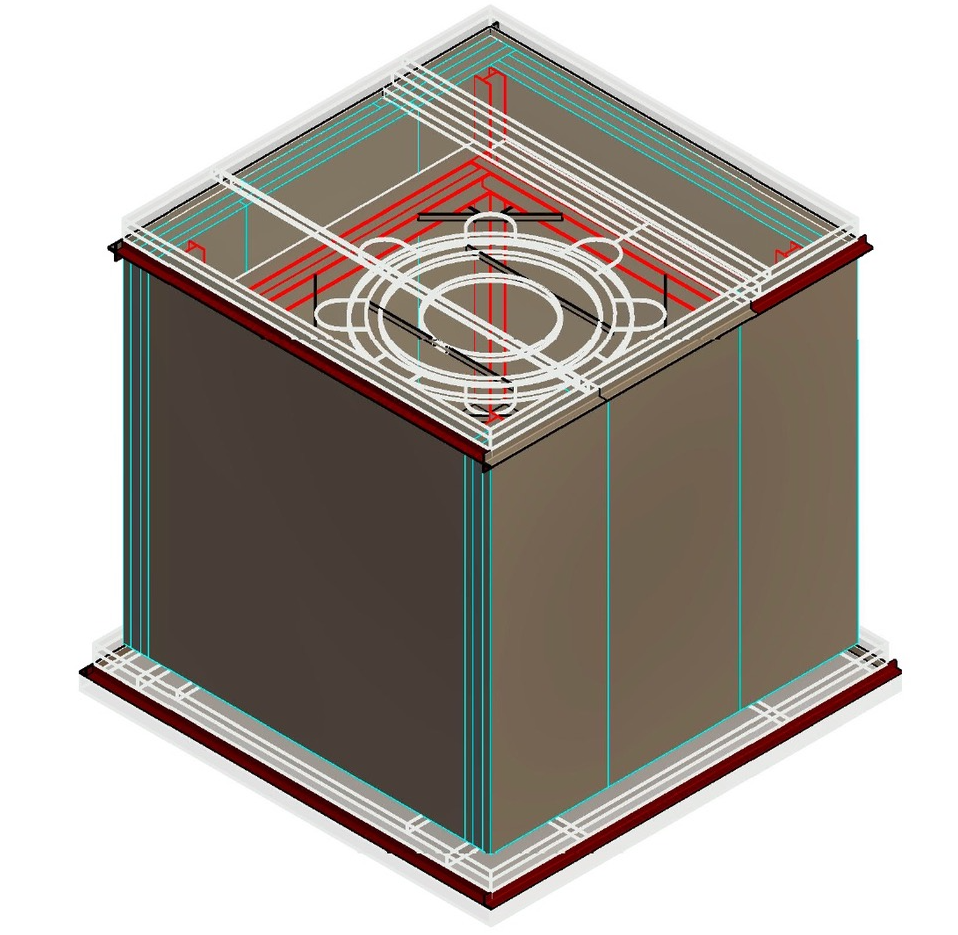}
\includegraphics[width=0.4\textwidth]{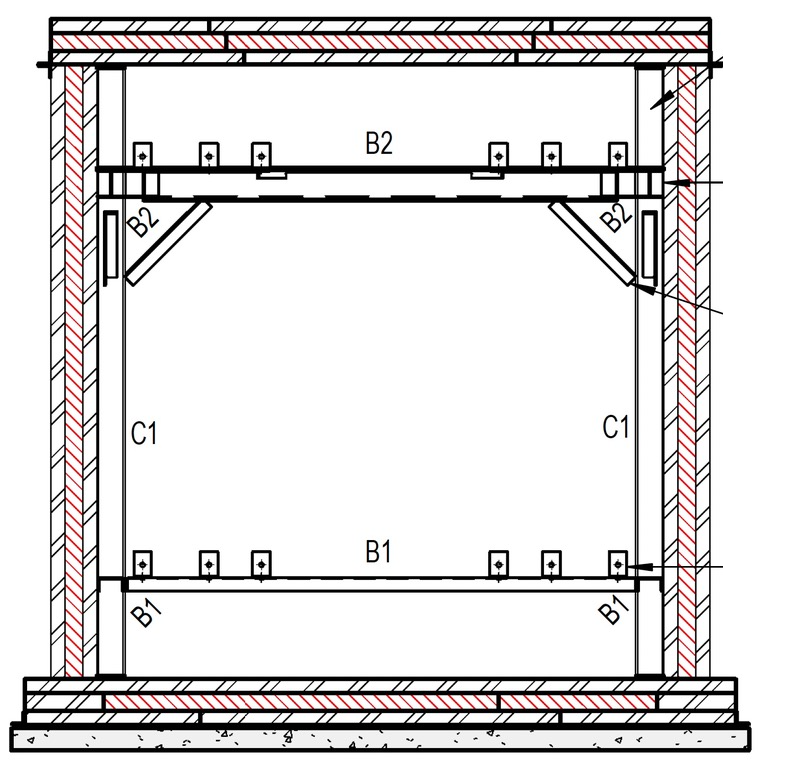}
  \caption{Left: Sketch of the radiation shielding with the internal support structure outline in red, and standing plate and roof section outline in white. Right: Cutaway of the shielding design, where the internal red layer represents that HDPE sandwiched by inner/outer layers of steel. The internal support frame is also depicted.}
  \label{fig:Shielding}
\end{figure}

Through simulations, steel shielding has been found to attenuate as much radiation as lead shielding with half the thickness. Interposing steel and high-density polyethylene (HDPE) provides a relatively compact and structurally sound shield, with the HDPE effective in neutron thermalisation and capture. Figure~\ref{fig:Shielding} provides a sketch of the shielding.

Other configurations of the shielding layers were considered, using alternating layers of gamma-ray attenuator (lead) and neutron attenuator (HDPE) keeping a constant total thickness of the two attenuators. The more gamma-ray attenuator there is in the innermost part of the shielding, the better the shielding is at attenuating gamma-rays and the worse it is at attenuating neutrons. For each configuration, the effect on the contribution of external radiation to the crystal background is just a few percent. The main difference between configurations is the rate of low-energy radiation, which is effectively absorbed by the liquid scintillator. The optimal solution is therefore a configuration that is structurally preferable. The final configuration is expected to bring the expected background rate from external radiation down to $\mathcal{O}$($10^{-4}$)~cpd/kg/keV$_{\rm ee}$.

Table~\ref{tab:shielding_activity} outlines the radiopurity of the shielding components. Samples of similar low-carbon aluminium-killed steel were screened by the Australian Nuclear Science and Technology Organisation (ANSTO) with a HPGe detector. No radioactive isotope was detected in the sample. This measurement was approximately ten times less sensitive than the measurements in Table~\ref{tab:bulk_steel}, and therefore we consider conservative limits. The radioactivity of the HDPE layers is not expected to contribute significantly to background, so HDPE samples were not measured. Instead, limits based on measurements from XENON are used~\cite{XENONScreening}.
\begin{table}[htbp!] 
\footnotesize
\centering
\begin{tabular}{ccc}
\hline
&  \multicolumn{2}{c}{Activity [mBq/kg]} \\
\cline{2-3}
Isotope & Steel & HDPE \\
\hline
$^{238}$U & $<13$ & 0.23\\
$^{232}$Th & $<6.7$ & $<0.14$\\
$^{40}$K & $<110$ & 0.7\\ 
$^{60}$Co & $<5.5$ & 0.06 \\ 
$^{137}$Cs & $<6.0$ & $<1.4$ \\ 
\hline
\end{tabular}
\caption{Radioactivity levels of the shielding steel and HDPE components~\cite{sabre_background}, based on steel radioactivity measured by ANSTO and HDPE values from XENON~\cite{XENONScreening}.
\label{tab:shielding_activity}}
\end{table}

\section{Expected sensitivity of SABRE South}
\label{sec:sensitivity}

Outlined below are the projected performance and sensitivity of the SABRE South detector, based on Geant4 simulation and a complete background model in the region of interest.

The key performance parameters are driven by the fact that, for NaI(Tl) crystals, a 1 keV deposition produces 30--45 photons, translating to 10--15 photoelectrons detected across the two PMTs in each module (based on a calculation using the light yield quoted in Table~\ref{tab:sabre-param}, and a PMT QE of 25--30\%). The SABRE South region of interest therefore has a series of distinct single-photon pulses distributed over several hundreds of nanoseconds. All currently operating NaI(Tl) experiments have background models that poorly reproduce the data at energies below 3 keV \cite{Coarasa:2024xec,BERNABEI2020,Adhikari}. Part of the difficulty in modelling the background in this low-energy region is due to the challenges of separating PMT noise, caused by spontaneous thermionic emission of electrons, from a genuine scintillation event of interest. Contributions from Cherenkov emissions in the glass of the crystal PMTs, and crystal afterglow following high intensity signals, are also being examined.

Table~\ref{tab:sabre-param}, lists the performance parameters relevant to the projected sensitivity. The quoted crystal light yield in this table is from measurements using PMTs with a QE lower than those in the final SABRE South detector. The light yield in SABRE South is expected to be approximately 20\% higher. 

\begin{table}[ht]
\centering
\caption{Key performance parameters of SABRE.}
\label{tab:sabre-param}
\begin{tabular}{@{}ll@{}}
\toprule
\bf{Parameter}                  & \bf{Value}           \\ \midrule
Crystal light yield                     & 11.1 $\pm$ 0.2 PE/keV \cite{antonello2021} \\
Crystal energy resolution               & 13.2\% (FWHM/E) at 59.5 keV \cite{antonello2021}               \\
Crystal energy threshold        & 1 keV               \\
Veto energy threshold           & 50 keV            \\
Total active mass               & 50 kg           \\
Background rate (South)         & 0.72 cpd/kg/keV \cite{sabre_background}                \\ \bottomrule
\end{tabular}
\end{table}

\begin{figure}[htb]
    \centering
    \includegraphics[height=0.48\textwidth]{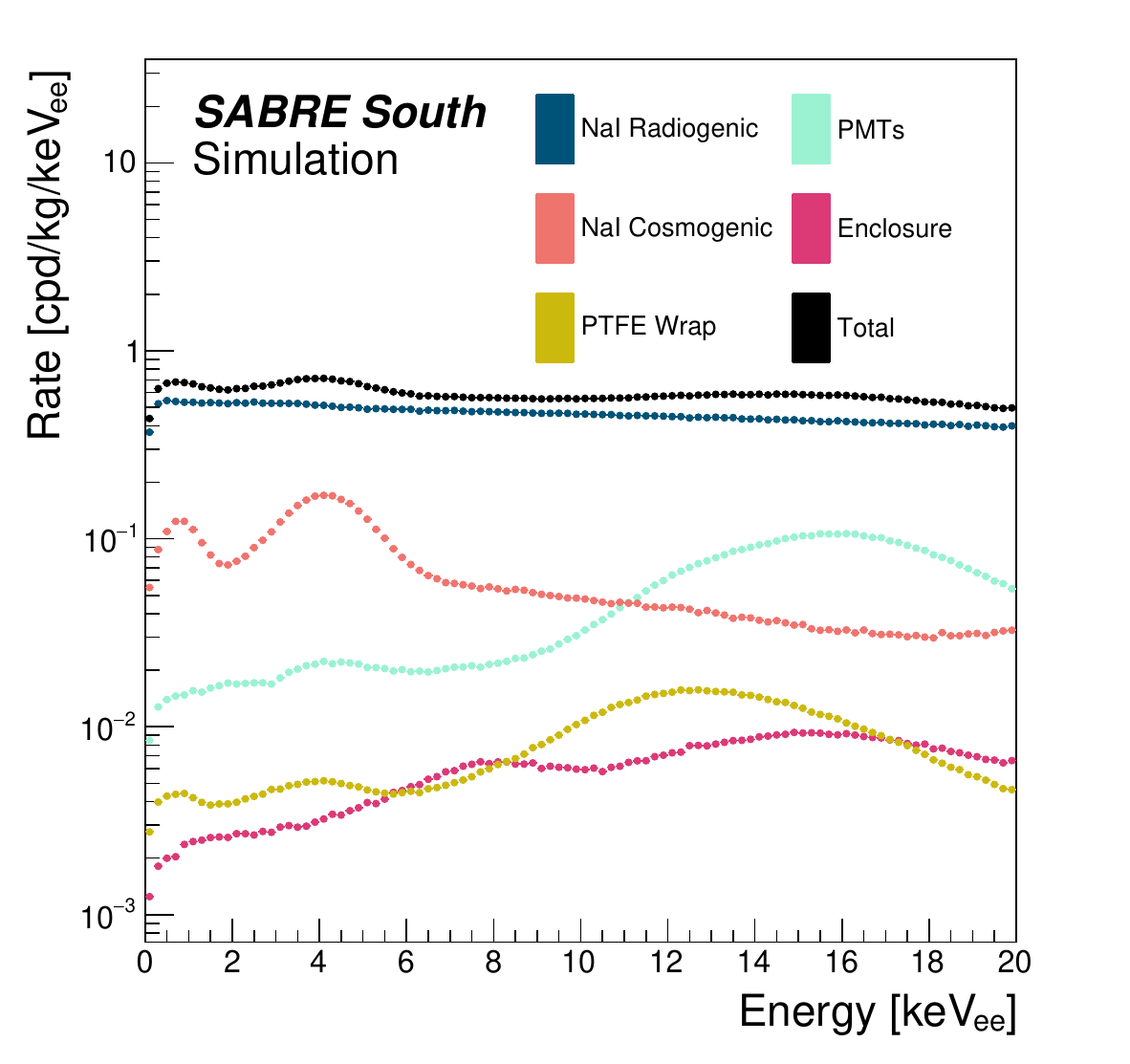}
    \includegraphics[height=0.44\textwidth]{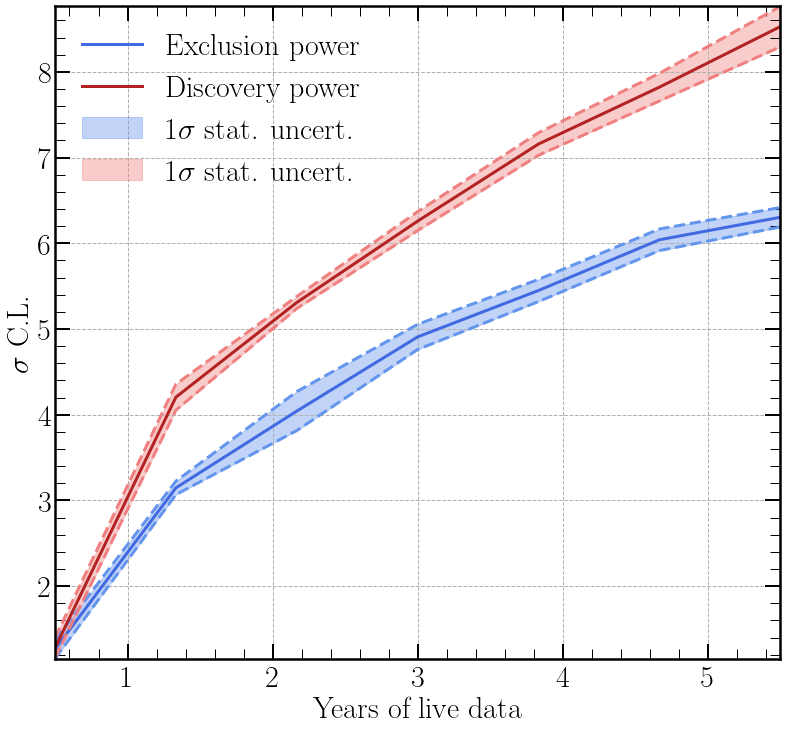}
    \caption{Left: Radioactive background as a function of energy deposited in the SABRE South crystals in the 0--20 keV range, with the total sum shown in black. A detailed description of the background contributions can be found in Ref.~\cite{sabre_background}. Right: The exclusion and discovery power of SABRE South for a DAMA-like signal, in the recoil energy region of 1--6~keV$_{\rm ee}$~\cite{sabre_background}. The shaded regions indicate 1$\sigma$ statistical uncertainty bands.}
    \label{fig:sensitivity}
\end{figure}

Assuming a total mass of 50 kg and a background of 0.72 cpd/kg/keV  (Fig.~\ref{fig:sensitivity}), the exclusion power and the discovery potential of SABRE South for the DAMA/LIBRA annual modulation signal \cite{Bernabei2018} are computed. The discovery potential of SABRE South to observe a signal with a modulation of 0.0118~cpd/kg/keV is calculated following the methodology of Ref.~\cite{Zurowski_2021} (Fig.~\ref{fig:sensitivity}) with a region of interest of 1--6~keV$_{\rm ee}$. This methodology involves producing distributions of $S_m$ (the modulation amplitude or rate of modulating events) for background only and signal plus background scenarios, where these distributions are generated by sampling from a background-only Poisson distribution for each time bin (assuming no modulation) or a signal plus background Poisson distribution (assuming a modulating signal). The discovery or exclusion power is determined based on the difference between the means of these distributions scaled by either $1/\sigma_{b}$ or $1/\sigma_{sb}$ respectively (where each $\sigma_{b,sb}$ derives from the $S_m$ distributions). With 3.1 annual cycles of data, SABRE South is able to refute the interpretation of the DAMA/LIBRA modulation as a dark matter signal with 5$\sigma$ CL. For observation of annual modulation, the modulation signal would reach a significance of 5$\sigma$ CL with two full years of data collection. This is possible because of the lower background rate compared to that of DAMA/LIBRA, and it is driven mainly by the development and optimisation of ultra-pure crystals and the active veto.

\section{Summary and outlook}
This document presents the executive summary of the technical design of the SABRE South detector, a NaI(Tl)-based detector that aims to test the long-standing DAMA/LIBRA annual modulation result. Using the same target material as the DAMA/LIBRA detector, the SABRE South NaI(Tl) crystals have an ultra-low background, owing to growth from Merck Astrograde NaI powder (which has potassium contamination below 10 ppb). SABRE South is also uniquely positioned in the Southern Hemisphere, allowing for potential disentanglement of seasonal effects. Crystal growth is ongoing through collaborations with SICCAS (China) and RMD (USA). Each crystal is coupled to two PMTs and enclosed in an oxygen-free, high-thermal-conductivity copper enclosure that is submerged in the SABRE South vessel (suspended by copper conduits). The vessel itself contains 11,600~L of liquid scintillator --- a mixture of LAB doped with approximately 3.5~g/L of PPO and 15~mg/L of bis-MSB. The vessel is instrumented with a baseline of 18 PMTs, and as many as 32. By including a liquid scintillator veto detector, and submerging the crystal detector modules within it, radioactive decays from within the crystals that may emit high-energy gamma rays can be tagged --- for example, $^{40}$K decays, which form a key impurity in the NaI(Tl) crystals. The liquid scintillator veto detector can also veto external background, with the possibility to employ reconstruction methods to disentangle background sources. In addition to the liquid scintillator veto, a 9.6 m$^2$ muon veto will be placed above the detector and shielding to veto incoming cosmogenic muons. This muon veto is also capable of position reconstruction of tagged muons along the length of the panels, and can be used in coincidence with the liquid scintillator veto.

The expected background rate for SABRE South in the 1--6~keV region is 0.72 cpd/kg/keV$_{\rm ee}$, which results in a 5$\sigma$ discovery (5$\sigma$ exclusion) power to a DAMA-like signal after 2 (3) years of data-taking. 
The SABRE South experiment is being installed in the Stawell Underground Physics Laboratory (SUPL) in 2024/2025, with the muon detectors having already been commissioned in telescope-mode as of early 2024.  

\section*{Acknowledgements}
The SABRE South program is supported by the Australian Government through the Australian Research Council (Grants: CE200100008, LE190100196, LE170100162, LE160100080, DP190103123, DP170101675, LP150100705). This research was partially supported by Australian Government Research Training Program Scholarships, and Melbourne Research Scholarships. This research was supported by The University of Melbourne's Research Computing Services and the Petascale Campus Initiative. We thank the SABRE North collaboration for their contributions to the SABRE South experiment design, in particular the crystal enclosure, the crystal insertion system and the gas handling system; and to the NaI(Tl) crystal development, background screening, and simulation framework. Work and equipment installation in the Stawell Underground Physics Laboratory is supported by the SUPL Board. We thank the Institute of High Energy Physics (IHEP), China, for their help in the procurement of the liquid scintillator, and for graciously providing us with the additional PMTs from the decommissioned Daya Bay experiment. For work on the crystal production, we thank RMD, Boston, and SICCAS, Shanghai, as well as Mellen (for the zone-refining process). For the production of the NaI-33 crystal, we thank Frank Calaprice and Jay Benziger from Princeton, alongside Seastar for providing the ICP-MS measurements, and Sandfire for providing the crucible in which the crystal was grown.

\newpage

\section*{List of Abbreviations}
    \textbf{ADC} Analogue-digital converter \\
    \textbf{ANSTO} Australian Nuclear Science and Technology Organisation \\
    \textbf{CPD} Counts Per Day \\
    \textbf{CPU} Central processing unit \\
    \textbf{CIS} Crystal insertion system \\
    \textbf{DAQ} Data acquisition \\
    \textbf{DM} Dark matter \\
    \textbf{EM} Electromagnetic \\
    \textbf{EPICS} Experimental Physics and Industrial Control System \\
    \textbf{EPMD} Ethylene propylene diene monomer \\
    \textbf{FPGA} Field Programmable Gate Array \\
    \textbf{HDPE} High-density polyethylene \\
    \textbf{HPGe} High purity Germanium \\
    \textbf{HPN} High purity nitrogen \\
    \textbf{HV} High voltage \\
    \textbf{IOC} Input-output controller \\
    \textbf{IPA} Isopropyl alcohol \\
    \textbf{LAB} Linear alkylbenzene \\
    \textbf{LNGS} Laboratori Nazionali del Gran Sasso \\
    \textbf{NI-cRIO} National Instruments compact reconfigurable input-output modules \\
    \textbf{OFHC} Oxygen-free, high-thermal-conductivity \\
    \textbf{PCB} Printed circuit board \\
    \textbf{PE} Photo electron \\
    \textbf{PLU} Programmable logic unit \\
    \textbf{PMMA} Polymethyl methacrylate \\
    \textbf{PMT} Photomultiplier tube \\
    \textbf{PTFE} Polytetrafluoroethylene \\
    \textbf{QE} Quantum efficiency \\
    \textbf{RAS} React Automation Studio \\
    \textbf{RCS} Radioisotope calibration system \\
    \textbf{RMD} Radiation Monitoring Devices \\
    \textbf{RMS} Root mean square \\
    \textbf{RTD} Resistance thermometer detector \\
    \textbf{SABRE} Sodium Iodide with Active Background REjection \\
    \textbf{SICCAS} Shanghai Institute Of Ceramics, Chinese Academy Of Sciences \\
    \textbf{SUPL} Stawell Underground Physics Laboratory \\
    \textbf{TTS} Transit time spread \\
    \textbf{VOC} Volatile organic compound \\
   
\clearpage
\bibliographystyle{JHEP}
\bibliography{refs.bib}

\end{document}